\pdfoutput=1
\documentclass[letterpaper]{article} % DO NOT CHANGE THIS
\usepackage{aaai2026}  % DO NOT CHANGE THIS
\usepackage{times}  % DO NOT CHANGE THIS
\usepackage{helvet}  % DO NOT CHANGE THIS
\usepackage{courier}  % DO NOT CHANGE THIS
\usepackage[hyphens]{url}  % DO NOT CHANGE THIS
\usepackage{graphicx} % DO NOT CHANGE THIS
\usepackage{natbib}  % DO NOT CHANGE THIS AND DO NOT ADD ANY OPTIONS TO IT
\usepackage{caption} % DO NOT CHANGE THIS AND DO NOT ADD ANY OPTIONS TO IT
\usepackage{algorithm}
\usepackage{algorithmic}
\usepackage[table,xcdraw]{xcolor}
\usepackage{tcolorbox}
\tcbuselibrary{breakable}

\usepackage{newfloat}
\usepackage{listings}
\DeclareCaptionStyle{ruled}{labelfont=normalfont,labelsep=colon,strut=off} % DO NOT CHANGE THIS
\floatstyle{ruled}
\newfloat{listing}{tb}{lst}{}
\floatname{listing}{Listing}
\title{SoMe: A Realistic Benchmark for LLM-based Social Media Agents}
\author {
    Dizhan Xue\textsuperscript{\rm 1,\rm 2},
    Jing Cui\textsuperscript{\rm 3},
    Shengsheng Qian\textsuperscript{\rm 1,\rm 2}\thanks{Corresponding Author.},
    Chuanrui Hu\textsuperscript{\rm 4},
    Changsheng Xu\textsuperscript{\rm 1,\rm 2,\rm 5}
}
\affiliations {
    \textsuperscript{\rm 1}State Key Laboratory of Multimodal Artificial Intelligence Systems, Institute of Automation, Chinese Academy of Sciences \\
    \textsuperscript{\rm 2}School of Artificial Intelligence, University of Chinese Academy of Sciences\\
    \textsuperscript{\rm 3}School of Computer Science and Engineering, Tianjin University of Technology\\
    \textsuperscript{\rm 4}Nanjing University of Posts and Telecommunications\\
    \textsuperscript{\rm 5}Peng Cheng Laboratory\\
    xuedizhan17@mails.ucas.ac.cn, cjtxdy@stud.tjut.edu.cn, shengsheng.qian@nlpr.ia.ac.cn, csxu@nlpr.ia.ac.cn
}

\usepackage{bibentry}
\begin{document}

\maketitle

\begin{abstract}
Intelligent agents powered by large language models (LLMs) have recently demonstrated impressive capabilities and gained increasing popularity on social media platforms.
While LLM agents are reshaping the ecology of social media, there exists a current gap in conducting a comprehensive evaluation of their ability to comprehend media content, understand user behaviors, and make intricate decisions.
To address this challenge, we introduce SoMe, a pioneering benchmark designed to evaluate social media agents equipped with various agent tools for accessing and analyzing social media data.
SoMe comprises a diverse collection of 8 social media agent tasks, 9,164,284 posts, 6,591 user profiles, and 25,686 reports from various social media platforms and external websites, with 17,869 meticulously annotated task queries.
Compared with the existing datasets and benchmarks for social media tasks, SoMe is the first to provide a versatile and realistic platform for LLM-based social media agents to handle diverse social media tasks.
By extensive quantitative and qualitative analysis, we provide the first overview insight into the performance of mainstream agentic LLMs in realistic social media environments and identify several limitations.
Our evaluation reveals that both the current closed-source and open-source LLMs cannot handle social media agent tasks satisfactorily.
SoMe provides a challenging yet meaningful testbed for future social media agents.
Our code and data are available at \url{https://github.com/LivXue/SoMe}.
\end{abstract}


\section{Introduction}

Language agents, which integrate tools with large language models (LLMs), have attracted broad research interest and demonstrate promising applications in human-level intelligent tasks, including coding \cite{zhang2024codeagent,islam2024mapcoder}, web browsing \cite{he2024webvoyager,li2025websailor}, and healthcare \cite{kim2024mdagents,li2024agent}.
Meanwhile, the development of social media agents, aimed at comprehending vast social data and executing human-like behaviors, has been a longstanding goal in the field of artificial intelligence \cite{edwards2014bot,carr2015social}.
Recent progress in LLM-based agents has also led to unprecedented prosperity of social media agents \cite{mou2024unveiling,zhang2025appagent}, which can potentially take on various tasks such as social event analysis \cite{qian2016multi}, post and user recommendation \cite{tang2014recommendation}, and social behavior simulation \cite{yang2024oasis}.
These agents can alleviate the burden on users and organizations managing an active online presence and analyzing vast amounts of social media data.
Moreover, these human-like agents are gradually active in the social networks of more and more human users.

\begin{figure*}[t]
\centering
\includegraphics[width=1.95\columnwidth]{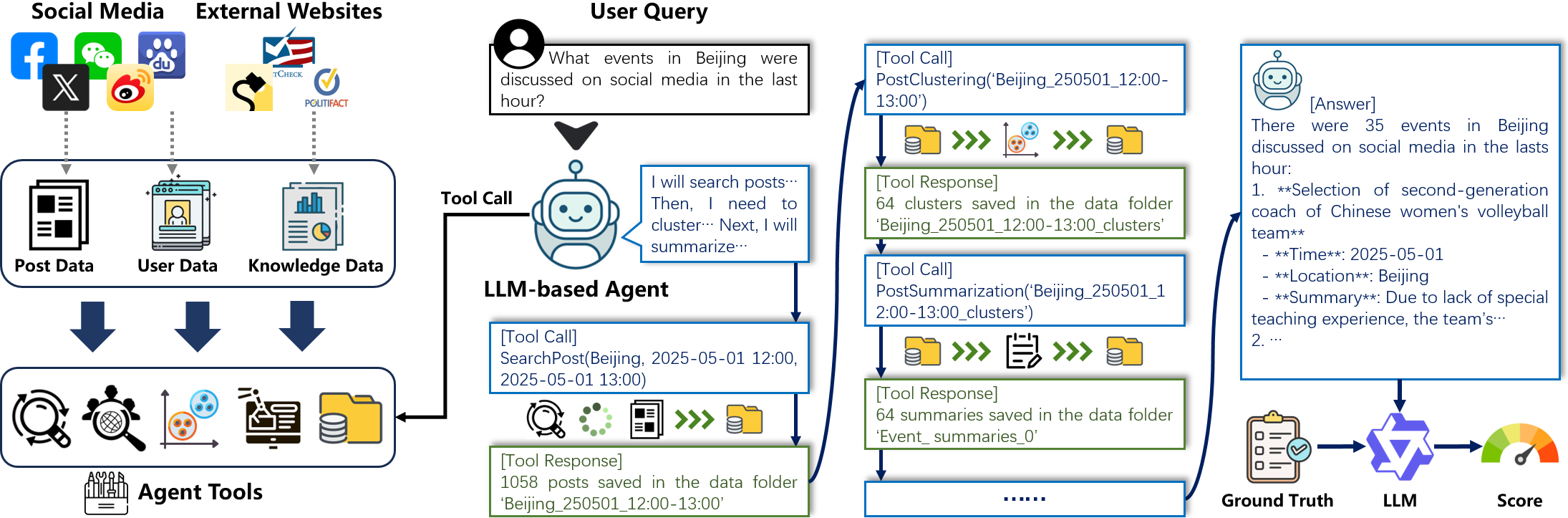}
\caption{\textbf{Workflow diagram of agents in SoMe.} The social media agent interacts with tools for data acquisition, management, and analysis, in order to arrive at an answer for the user query. The answer is evaluated with the assistance of an LLM scorer.}
\label{fig1}
\end{figure*}

While the increasing participation of social media agents is reshaping the ecology of social media, the incomplete understanding of their capability in multiplexed social media tasks has raised concerns.
Previous evaluations of agents and LLMs in the area of social media typically focus on a single task, such as misinformation detection \cite{nakazato2024jsocialfact} or user behavior prediction \cite{jiang2023social}.
For example, BotSim \cite{qiao2025botsim} designs a user behavior simulation framework for LLM-based agents, powered by tens of thousands of data collected from Reddit.
TrendSim \cite{zhang2025trendsim} creates a user simulation environment for social media agents that incorporates a time-aware interaction mechanism, based on data of 1,000 users collected from Weibo. 
These existing evaluations are also limited by insufficient data and the lack of ground truth (e.g., TrendSim utilizes LLMs to evaluate the rationality of agent behaviors without ground truth references).
To sum up, these evaluations are insufficient for providing a full range of knowledge about social media agents.
To further catalyze the research, a versatile benchmark with abundant real data and annotations is required for developing and deploying trustworthy social media agents.

To address the abovementioned challenges, we propose SoMe, the first versatile benchmark for social media agents, which aims at comprehensively evaluating the agentic capabilities of LLMs on social media tasks.
SoMe assesses LLMs across 8 critical tasks: real-time event detection, streaming event summarization, misinformation detection, user behaviour prediction, user emotion analysis, user comment simulation, media content recommendation, and social media question-answering.
To enable LLM-based agents to perform these tasks, we build a platform with 8 agent tools for acquiring, managing, and analyzing data in and outside of social media.
Moreover, we collect 9 million of real and public data from 32 social media platforms (e.g., X\footnote{https://x.com} and Weibo\footnote{https://weibo.com/}) and external websites, aligning with the practical application of social media agents. 
These data cover a wide range of topics, events, and users, while remaining real and challenging attributions of social media data, which are noisy, temporal, and diverse.
We leverage human-LLM interactive pipelines to annotate task queries and their ground truth for all tasks, elaborately verified by 10 professional annotators in total.
In our benchmark, LLM-based agents are required to conduct step-by-step data processing and long-context reasoning with appropriate tool calls in a realistic environment, as exemplified in Figure \ref{fig1}.
Brief statistics of data are demonstrated in Table \ref{tab:data}.

\begin{table}[t]
\scriptsize
\centering
\setlength{\tabcolsep}{1mm} 
\begin{tabular}{c|ccc}
\hline
\textbf{Task}                   & \textbf{\#Query}     & \textbf{\#Data}      & \textbf{Data Type}   \\ \hline
Real-time Event Detection       & 568                  & 476,611              & Posts                \\
Streaming Event Summarization   & 154                  & 7,898,959            & Posts                \\
Misinformation Detection        & 1,451                & 27,137               & Posts \& Knowledge   \\
User Behavior Prediction       & 3,000                & 840,200              & Posts \& Users       \\
User Emotion Analysis           & 2,696                & 840,200              & Posts \& Users       \\
User Comment Simulation         & 4,000                & 840,200              & Posts \& Users       \\
Media Content Recommendation    & 4,000                & 840,200              & Posts \& Users       \\
Social Media Question-answering & 2,000                & 8,651,759            & Posts \& Users       \\ \hline
Total                           & 17,869               & 9,242,907            & All \\ \hline
\end{tabular}
\caption{Statistics of data in SoMe: \textbf{\#Query} denotes the number of queries (with annotations) in the dataset, \#Data denotes the number of accessible data in the database, and \textbf{Data Type} denotes the type of data in the database.}
\label{tab:data}
\end{table}

Our contributions are summarized as follows:
\begin{itemize}
    \item We propose SoMe, the first realistic evaluation benchmark for LLM-based social media agents. SoMe comprises 8 tasks and millions of real data, assessing capabilities across social data analysis, user personality understanding, and long-context knowledge reasoning.
    \item We establish a running platform for LLM-based social media agents with 8 tools for data acquisition, management, and analysis. The agents can perform complex tasks in social media by step-by-step data processing and reasoning with appropriate tool combinations.
    \item We provide the overview insight into the performance of 13 mainstream agentic LLMs in realistic social media environments. Our findings reflect the bottleneck of existing social media agents, providing suggestions for future work.
\end{itemize}

\section{Related Work}
\subsection{LLM-based Social Media Agents}
Social media has emerged as one of the most popular technological innovations globally, attracting billions of users and fundamentally reshaping how individuals, organizations, and societies communicate, interact, and exchange information. 
Due to its tremendous social impact, the research on social media has become an active and long-standing area in artificial intelligence \cite{roy2012socialtransfer,qian2016multi,qian2023open,xue2025short}.
However, the open-world nature and sophisticated structure of social media significantly reduce the effectiveness of models with static inputs.
Recently, in the pursuit of developing intelligent agents, there has been considerable
focus on integrating LLMs with external tools \cite{wang2024badagent,xue2024few,he2025plan,qian2024linguistic,yangsvbench}.
Specially, agentic LLMs \cite{hurst2024gpt,comanici2025gemini,yang2025qwen3} are proposed to enhance the abilities of using external tools and handling agentic tasks.
Based on agentic LLMs, these agents enable powerful capabilities in environment interaction, decision-making, and task execution.
Especially, LLM-based social media agents exhibit unprecedented ability in multi-round information acquisition, analysis, and reasoning, boosting performance in tasks such as misinformation detection \cite{wan2024dell}, user simulation \cite{gao2024simulating}, and event analysis \cite{wang2024news}.
For instance, \citet{wang2024news} design an LLM-based agent to iteratively filter out irrelevant news and employ human-like
reasoning to analyze complex social events.
\citet{zhang2025trendsim} propose an LLM-based multi-agent system to simulate user interactions in trending topics on social media. 
\citet{wei2024mimicking} devise anonymous opinion leader agents to simulate and predict the user emotional responses to different events on social media.

However, these social media agents specialize in different tasks for social media and cannot handle multiple tasks simultaneously. To thoroughly evaluate the capabilities of social media agents, we propose SoMe to measure the performance across 8 agentic social media tasks with a versatile agent platform. This platform is equipped with various tools enabling interactions within a realistic social media environment.

\subsection{Benchmarks for LLM-based Agents}
With the rise of LLM-based agents, the performance and trustworthiness of agents in various real-world tasks are of concern \cite{liuagentbench,he2025plan}.
Numerous studies have been conducted to evaluate the tool-use and environment-interaction abilities of LLM-based agents \cite{xie2024osworld,skrynnikpogema}.
For example, CharacterEval \cite{tu-etal-2024-charactereval} evaluates the role-playing ability of agents in multi-turn role-playing dialogues.
OSWorld \cite{xie2024osworld} evaluates the real-world computer-using ability of multimodal agents in a real computer environment.
VisualWebArena \cite{koh-etal-2024-visualwebarena} assesses the instruction execution capability of multimodal web agents on realistic visually grounded tasks.
While existing benchmarks have evaluated LLM-based agents in various application domains, a comprehensive benchmark for social media agents is still lacking, despite their prevalence in real-world applications.

Unlike environments with explicit guidance in the above benchmarks, social media is an extremely noisy environment, where information about a concept can usually be scarce or inundated with massive irrelevant information \cite{baldwin2013noisy}.
Therefore, we establish SoMe, a novel and comprehensive benchmark that aims to address this gap by offering an elaborate and realistic evaluation framework for LLM-based social media agents.

\section{SoMe Benchmark}
In this section, we will introduce the task definition of social media agents, the collected data, and the supported tools of the constructed evaluation platform.

\subsection{Social Media Tasks}
Evaluating the capabilities of social media agents necessitates a comprehensive and structured approach. We collect the 8 most-considered agentic tasks related to social media, categorized into 3 classes, as follows:
\begin{itemize}
    \item \textbf{Post-centered Tasks:} This class of tasks requires multi-round data processing and interactions with social media posts, sometimes necessitating access to external knowledge bases. \textbf{Real-time Event Detection (RED)} aims to detect social events in real-time from a large volume of recent posts. \textbf{Streaming Event Summarization (SES)} aims to progressively summarize the details of a social event from continuously published posts. \textbf{MisInformation Detection (MID)} aims to identify false, misleading, or inaccurate information within posts, utilizing support from external knowledge.

    \item \textbf{User-centered Tasks:} This class of tasks involves understanding user preferences and behavior patterns through multi-round tool calls, thereby making personalized predictions. \textbf{User Behavior Prediction (UBP)} aims to predict the user interaction behaviors with specific posts. \textbf{User Emotion Analysis (UEA)} aims to predict the emotions that emerge from users towards particular posts. \textbf{User Comment Simulation (UCS)} aims to predict the comments that users will make on given posts.

    \item \textbf{Comprehensive Tasks}: This class of tasks needs comprehensive analyses of both extensive posts and users to deduce the answers. \textbf{Media Content Recommendation (MCR)} aims to recommend social media content that aligns with user preferences. \textbf{Social Media Question-answering (SMQ)} aims to answer questions regarding the public information available in posts and users.
\end{itemize}
Details and examples of these tasks are introduced in Appendix.

\subsection{Data Sources and Annotations}
\textbf{Data sources.} We collect a large amount of data from primarily 32 social media platforms, such as X and Weibo, with a predominant focus on English and Chinese content.
Specifically, these data are from three sources: 1) We buy commercial data from the Xiaoying company\footnote{https://www.xiaoying.tv/}, which are publicly visible data and crawled from diverse social media platforms. 2) We crawl data of top influencers and users in their social networks on Weibo. 3) We adopt open-source datasets \cite{yang2022coarse}, where the fact-checked reports from external websites are separated to form a knowledge base for misinformation detection.
To sum up, there are 9,164,284 posts, 6,591 user profiles, and 25,686 reports in our database, which can be accessed through various agent tools.

\textbf{Data annotations.} While text is the major modality in social media, we convert images and videos in posts into captions (by Qwen2.5-VL-72B \cite{bai2025qwen2}), OCR texts (by Qwen2.5-VL-72B), and ASR texts (by GPT-4o \cite{hurst2024gpt}).
After that, we propose a semi-automated annotation pipeline for all tasks.
Specifically, for UBP, UCS, and MCR, we automatically generate task queries about the collected users and convert the collected data into ground truth annotations based on templates.
Since the results are already included in the collected data, no LLM- or human-involved edition of ground truth is involved for these tasks.
For RED, SES, UEA, and SMQ, we leverage multi-round human-LLM interactive pipelines to annotate ground truth results for automatically generated task queries, based on Qwen3-32B.
All ground truth results undergo manual filtering and verification by 10 professional annotators to guarantee the quality of annotations. 
For MID, we merge the open-source LIAR-RAW and RAWFC datasets \cite{yang2022coarse}, where the ground truth is already annotated.

More details of our data sources and annotations are included in Appendix.

\begin{figure*}[ht]  
\centering
    \begin{minipage}{0.32\linewidth}
        \centerline{\includegraphics[width=\linewidth]{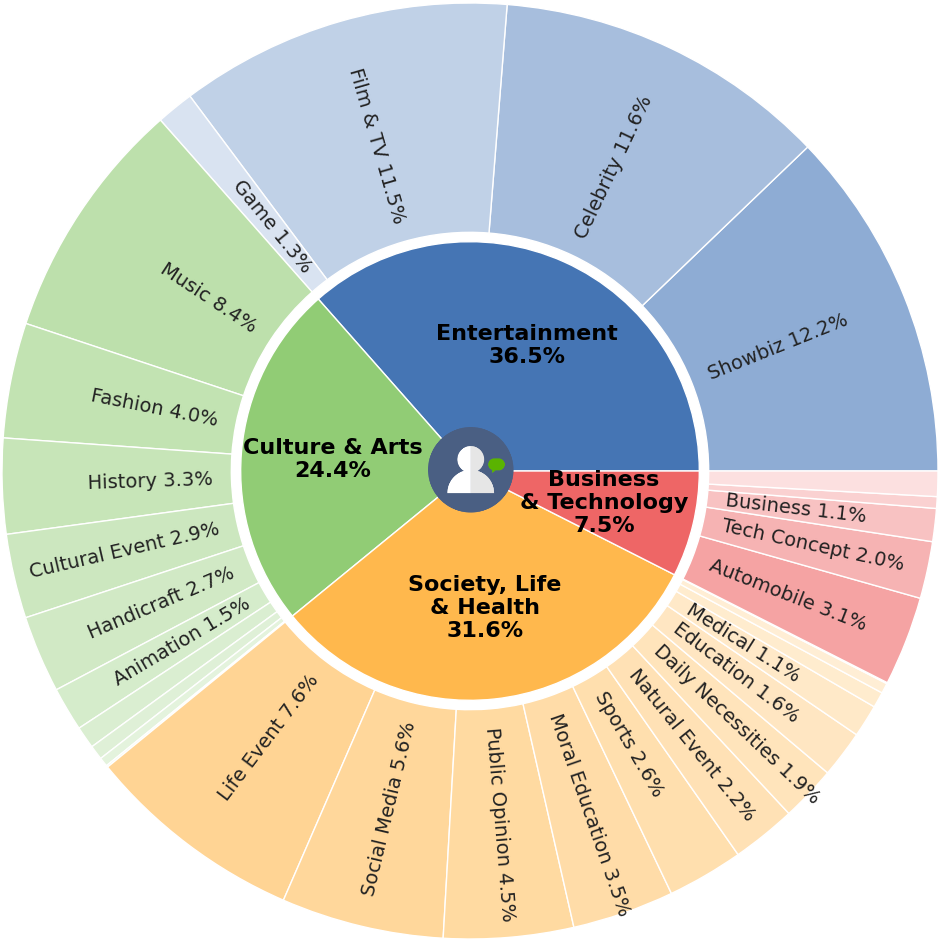}}
        \centerline{\footnotesize{(a) User interest distribution}}
    \end{minipage}
    \hfill
    \begin{minipage}{0.32\linewidth}
        \centerline{\includegraphics[width=\linewidth]{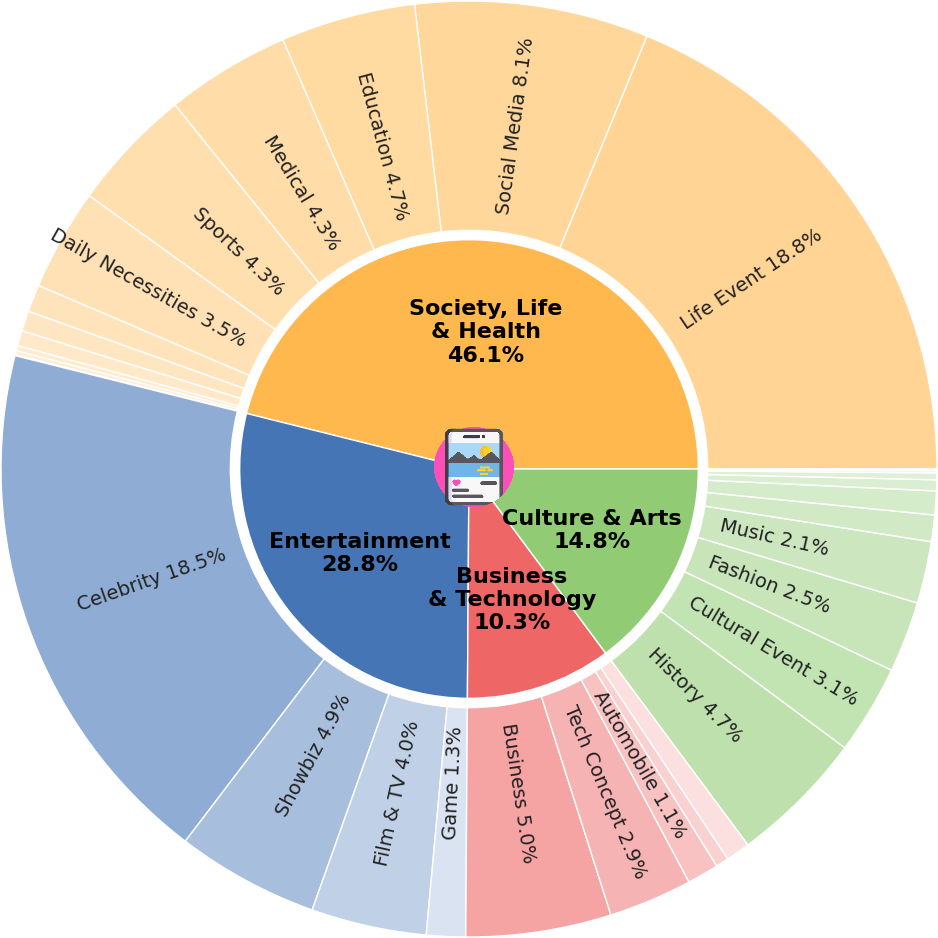}}
        \centerline{\footnotesize{(b) Post topic distribution}}
    \end{minipage}  
    \hfill
    \begin{minipage}{0.32\linewidth}
        \centerline{\includegraphics[width=\linewidth]{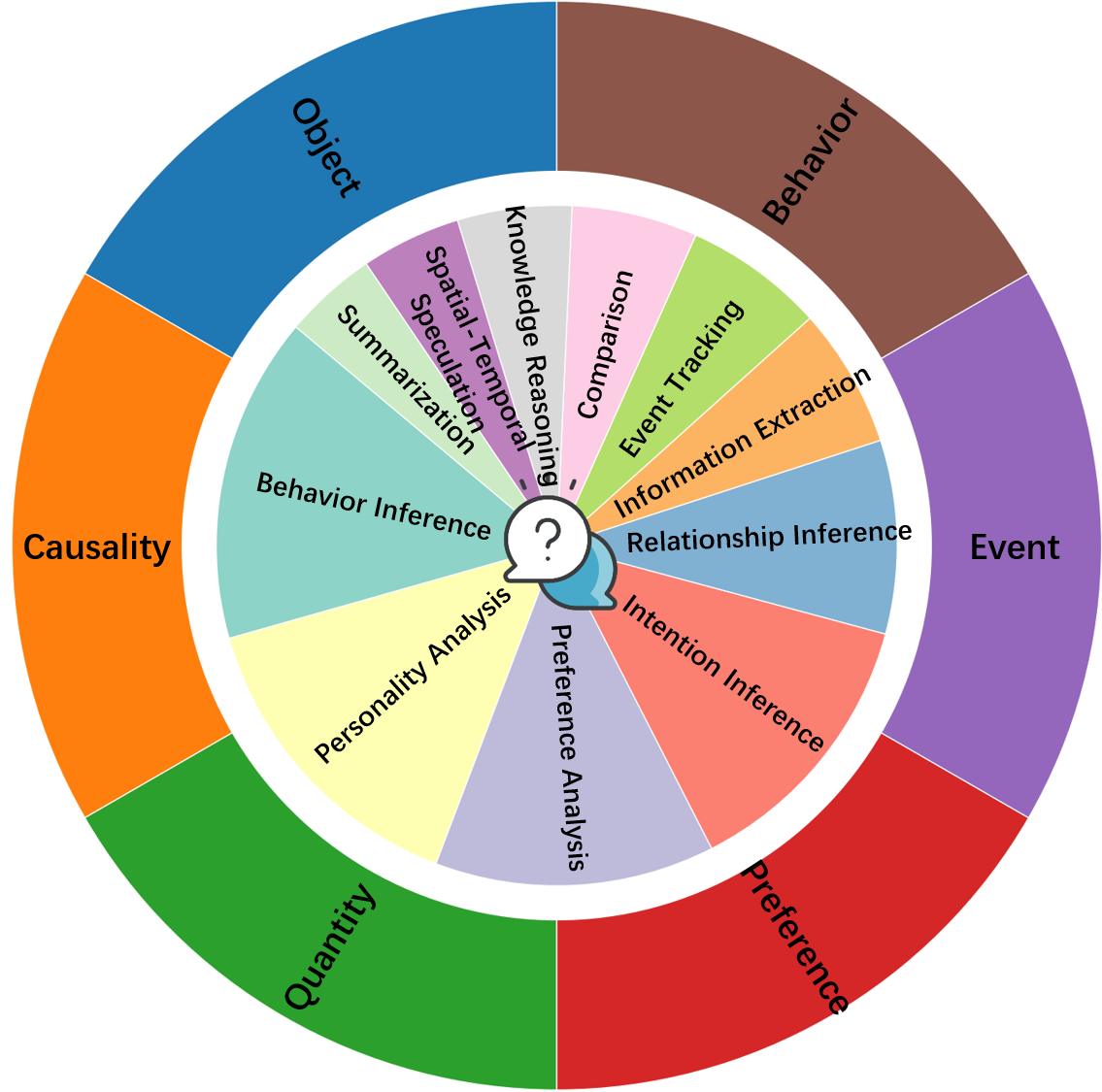}}
        \centerline{\footnotesize{(c) Task query distribution}}
    \end{minipage}  
    \vfill
    \caption{Distributions of social media data in SoMe.}
    \label{fig:dis}
\end{figure*} 

\begin{table*}[ht]
\footnotesize
\centering
\begin{tabular}{l|l}
\hline
\textbf{Tool Name(Parameters)}                & \textbf{Description}                                                                  \\ \hline
DataFolder(folder\_name, start\_idx, end\_idx) & Output data in the data folder of a given name, from a start index to an end index.   \\
SearchPost(location, start\_time, end\_time)  & Search posts about a location published in a time period, save them in a data folder. \\
SearchTopic(topic\_name)                       & Search posts about a given topic and save them in a data folder.                      \\
SearchUser(uid)                                & Search a specific user, output the profile, and save the posts into a data folder.    \\
RetrievePost(query, folder\_name, topk)       & Retrieve the top-k most relevant posts with the query in a data folder.               \\
RetrieveKnowledge(query, topk)                 & Retrieve the top-k most relevant reports with the query in the knowledge base.        \\
PostClustering(folder\_name)                   & Cluster posts in a data folder and save clusters in another data folder.              \\
PostSummarization(folder\_name)                & Summarize post clusters in a data folder and save summaries in another data folder.   \\ \hline
\end{tabular}
\caption{Agent tools in SoMe.}
\label{tab:tool}
\end{table*}

\subsection{Data statistics}
The brief statistics of the dataset are shown in Table \ref{tab:data}.
We conduct further investigation on the data diversity of SoMe.

\textbf{User analysis.} The user interests identified by Qwen3-32B based on the collected user profiles are presented in Figure \ref{fig:dis}(a), showcasing the outline and a broad spectrum of user interests.  
Specifically, there are 4 major interests, including Entertainment (36.5\%), Society, Life, \& Health (31.6\%), Culture \& Arts (24.4\%), and Business \& Technology (7.5\%), and 33 subcategories.
Among subcategories, Showbiz (12.2\%), Celebrity (11.6\%), Film \& TV (11.5\%), Music (8.4\%), and Life Event (7.6\%) are the most popular.
Overall, the interests of users in our data are highly diverse, as the most popular one only accounts for 12.2\%, which improves the comprehensiveness and robustness of our benchmark.

\textbf{Post analysis.} 
The topic distribution of the collected posts, as identified by Qwen3-32B, is presented in Figure \ref{fig:dis}(b). The results reveal a diverse spread across four primary domains aligned with user interests: Society, Life \& Health (46.1\%), Entertainment (28.8\%), Culture \& Arts (14.8\%), and Business \& Technology (10.3\%). Within these categories, Life Events (18.8\%) and Celebrities (18.5\%) emerged as the most prevalent subtopics. Significantly, all other identified subtopics individually accounted for less than 10\% of the posts. 
This substantial heterogeneity within the collected social media posts underscores the breadth of content captured. 
Consequently, this diverse representation bolsters both the comprehensiveness and the robustness of our benchmark.

\textbf{Query analysis.} 
We further investigate the queries in all 8 tasks, as shown in Figure \ref{fig:dis}(c).
Task queries in SoMe can be roughly classified into questions about behaviors, events, preferences, quantity, causality, and objects.
Moreover, we classify all queries into 11 categories, of which the percentages are presented in the figure.
Each category corresponds to the assessment for one specific skill of social media agents.
The criteria for these categories are as follows: 
(1) \textit{Behavior Inference}: Analyzing the user behavior patterns and predicting user behaviors. 
(2) \textit{Personality Analysis}: Identifying the personality of users based on their social media activities.
(3) \textit{Preference Analysis}: Inferring and interpreting the preferences of users.
(4) \textit{Intent Inference}: Discerning the underlying intention behind actions and speech.
(5) \textit{Relationship Inference}: Discovering the relationships between people, objects, or events.
(6) \textit{Information Extraction}: Extracting key information from noisy data for specific queries.
(7) \textit{Event Tracking}: Tracking the processes and transitions of events within long contexts.
(8) \textit{Comparison}: Comparing the information about multiple objects or from multiple sources.
(9) \textit{Knowledge Reasoning}: Retrieving and applying external knowledge to conduct logical reasoning.
(10) \textit{Spatial-Temporal Speculation}: Understanding the spatial and temporal relationships within social media data.
(11) \textit{Summarization}: Integrating information from various sources to provide thorough and succinct reports.

\begin{table*}[t]
\centering
\begin{tabular}{lcccccccccc}
\hline
\multicolumn{1}{l|}{\textbf{Model}}            & \multicolumn{1}{c|}{\textbf{Size}} & \textbf{RED}         & \textbf{SES}         & \textbf{MID}         & \textbf{UBP}         & \textbf{UEA}         & \textbf{UCS}         & \textbf{MCR}         & \multicolumn{1}{c|}{\textbf{SMQ}}         & \textbf{Avg.}         \\ \hline
\multicolumn{11}{l}{\cellcolor[HTML]{96FFFB}\textbf{API-based}}                                                                                                                                                                                                             \\ 
\multicolumn{1}{l|}{GPT-4o}                    & \multicolumn{1}{c|}{N/A}           & 47.59                     & 36.17                     & 50.24                     & 55.17                     & 31.53                     & 52.48                     & 61.63                     & \multicolumn{1}{c|}{64.21}                    & 49.88                     \\
\multicolumn{1}{l|}{Gemini-2.5-Flash}          & \multicolumn{1}{c|}{N/A}           & \textbf{54.92}                     & \textbf{44.87}                     & 45.62                     & 57.50                     & 41.94                     & 56.00                     & 62.75                     & \multicolumn{1}{c|}{71.01}                    & \textbf{54.33}                     \\
\multicolumn{1}{l|}{Kimi-K2-Instruct}          & \multicolumn{1}{c|}{1T}            & 50.40                     & 38.57                     & 47.83                     & 51.50                     & \textbf{45.94}                     & \textbf{58.25}                     & 57.00                     & \multicolumn{1}{c|}{77.38}                    & 53.36                     \\
\multicolumn{1}{l|}{DeepSeek-V3}               & \multicolumn{1}{c|}{671B}          & 35.94                     & 38.83                     & \textbf{51.00}                     & 55.06                     & 42.98                     & 54.92                     & 56.25                     & \multicolumn{1}{c|}{75.94}                    & 51.37                     \\ \hline
\multicolumn{11}{l}{\cellcolor[HTML]{9AFF99}\textbf{Open-source}}                                                                                                                                                                                                           \\ 
\multicolumn{1}{l|}{Llama-3.3-70B-Instruct}    & \multicolumn{1}{c|}{70B}           & 34.84                     & 33.77                     & 50.24                     & 55.83                     & 32.77                     & 53.00                     & 61.40                     & \multicolumn{1}{c|}{64.83}                    & 48.34                     \\
\multicolumn{1}{l|}{Qwen3-32B}                 & \multicolumn{1}{c|}{32B}           & 44.25                     & 41.04                     & 47.42                     & \textbf{67.03}                     & 33.53                     & 54.98                     & \textbf{63.28}                     & \multicolumn{1}{c|}{\textbf{80.27}}                    & 53.98                     \\
\multicolumn{1}{l|}{GLM-4-32B-0414}            & \multicolumn{1}{c|}{32B}           & 27.55                     & 27.27                     & 41.38                     & 40.17                     & 26.10                     & 47.13                     & 44.08                     & \multicolumn{1}{c|}{57.19}                    & 38.86                     \\
\multicolumn{1}{l|}{Devstral-Small-2507}       & \multicolumn{1}{c|}{24B}           & 30.99                     & 33.31                     & 47.07                     & 47.93                     & 22.43                     & 46.68                     & 47.75                     & \multicolumn{1}{c|}{58.25}                    & 41.80                     \\
\multicolumn{1}{l|}{Qwen3-14B}                 & \multicolumn{1}{c|}{14B}           & 43.97                     & 40.19                     & 44.80                     & 66.20                     & 33.35                     & 54.43                     & 62.13                     & \multicolumn{1}{c|}{77.47}                    & 52.82                     \\
\multicolumn{1}{l|}{GLM-4-9B-0414}             & \multicolumn{1}{c|}{9B}            & 14.42                     & 29.68                     & 33.63                     & 48.73                     & 28.63                     & 44.10                     & 49.05                     & \multicolumn{1}{c|}{51.82}                    & 37.51                     \\
\multicolumn{1}{l|}{Qwen3-8B}                  & \multicolumn{1}{c|}{8B}            & 40.38                     & 36.69                     & 45.21                     & 61.73                     & 33.03                     & 53.33                     & 60.55                     & \multicolumn{1}{c|}{76.18}                    & 50.89                     \\
\multicolumn{1}{l|}{DeepSeek-R1-0528-Qwen3-8B} & \multicolumn{1}{c|}{8B}            & 17.71                     & 28.83                     & \underline{26.46}                     & 43.53                     & \underline{21.18}                     & \underline{31.10}                     & \underline{34.33}                     & \multicolumn{1}{c|}{51.84}                    & \underline{31.87}                     \\
\multicolumn{1}{l|}{Llama-3.1-8B-Instruct} & \multicolumn{1}{c|}{8B}                & \underline{3.37}                     & \underline{20.78}                     & 40.45                     & \underline{34.23}                     & 37.11                     & 33.98                     & 47.40                     & \multicolumn{1}{c|}{\underline{31.05}}                    & 31.65                     \\ \hline
\end{tabular}
\caption{Performance of agentic LLMs evaluated on 8 tasks in SoMe. Avg. denotes the average scores over 8 tasks. The highest are highlighted in bold while the lowest are marked with underlines.}
\label{tab:res}
\end{table*}

\subsection{Agent Tools}
We construct a versatile framework for social media agents with 8 tools for data acquisition, management, and analysis. All tools are designed in accordance with Model Context Protocol (MCP) \cite{hou2025model} to ensure broad compatibility with mainstream agentic LLMs.
These tools are summarized in Table \ref{tab:tool}.
Detailed descriptions of all tools and their parameters are provided to agents following MCP, ensuring a full understanding of the tools.
Finally, upon receiving the task query, the constructed social media agents can interact with appropriate tools and conduct step-by-step reasoning to handle the task.

\section{Evaluation}
More details of the evaluation and experiments are included in Appendix.
\subsection{Experiment Settings}
We evaluate 13 mainstream agentic LLMs on SoMe. For API-based models, we select GPT-4o \cite{hurst2024gpt}, Gemini-2.5-Flash \cite{comanici2025gemini}, Kimi-K2-Instruct \cite{kimik2}, and DeepSeek-V3 \cite{liu2024DeepSeek}. For open-source models, we adopt Qwen3 series \cite{yang2025qwen3}, Llama-3 series \cite{dubey2024llama}, GLM-4 series \cite{glm2024chatglm}, Devstral-Small-2507 \cite{jiang2023mistral7b}, and DeepSeek-R1-0528-Qwen3-8B \cite{guo2025DeepSeek}.
Though Kimi-K2-Instruct and DeepSeek-V3 are also open-source, we opt for API calling due to the overly large size of these models.
Experiments are conducted using 8 NVIDIA A800-SMX4-80GB GPUs with the vLLM framework \cite{kwon2023efficient} for open-source model deployment.
All LLMs are equipped with agent tools using the standard Model Context
Protocol (MCP), while the native tool-calling formats of all LLMs are supported and correctly parsed in our agent framework.

\subsection{Evaluation Metrics}
We separately design appropriate metrics for 8 social media agent tasks.
For RED, SES, and SMQ, we employ LLM-based metrics to compare the semantics of the generated results with the ground truth and assign corresponding scores.
For MID, UBP, UEA, UCS, and MCR, we utilize LLMs to filter out redundant information and extract the key answers in the agent responses.
Then, we compute the accuracy (ACC) of the answers based on the ground truth.
Additionally, we compute the Task Completion Rate (TCR) for all tasks, which measures the percentage of task queries an agent successfully completes in a specific task.
All metric scores are normalized to the range of $[0, 100]$, and we report the average scores over test samples for all tasks.
Details of the metrics and LLM prompts are included in Appendix.

\begin{table*}[t]
\centering
\begin{tabular}{lcccccccccc}
\hline
\multicolumn{1}{l|}{\textbf{Model}}            & \multicolumn{1}{c|}{\textbf{Size}} & \textbf{RED}         & \textbf{SES}         & \textbf{MID}         & \textbf{UBP}         & \textbf{UEA}         & \textbf{UCS}         & \textbf{MCR}         & \multicolumn{1}{c|}{\textbf{SMQ}}         & \textbf{Avg.}         \\ \hline
\multicolumn{11}{l}{\cellcolor[HTML]{96FFFB}\textbf{API-based}}                                                                                                                                                                                                             \\ 
\multicolumn{1}{l|}{GPT-4o}                    & \multicolumn{1}{c|}{N/A}           & 94.54                     & 90.91                     & 95.11                     & 98.00                     & 98.70                     & 96.45                     & 97.35                     & \multicolumn{1}{c|}{87.50}                    & 94.82                     \\
\multicolumn{1}{l|}{Gemini-2.5-Flash}          & \multicolumn{1}{c|}{N/A}           & \cellcolor[HTML]{DBFFDB}99.23                     & 94.81                     & 88.63                     & 95.50                     & \cellcolor[HTML]{DBFFDB}99.04                     & \cellcolor[HTML]{DBFFDB}99.75                     & \cellcolor[HTML]{DBFFDB}99.25                     & \multicolumn{1}{c|}{96.63}                    & 96.61                     \\
\multicolumn{1}{l|}{Kimi-K2-Instruct}          & \multicolumn{1}{c|}{1T}            & 98.86                     & 85.06                     & 94.21                     & 97.17                     & 98.87                     & \cellcolor[HTML]{DBFFDB}99.75                     & \cellcolor[HTML]{DBFFDB}99.25                     & \multicolumn{1}{c|}{98.03}                    & 96.40                     \\
\multicolumn{1}{l|}{DeepSeek-V3}               & \multicolumn{1}{c|}{671B}          & \cellcolor[HTML]{DBFFDB}99.36                     & 92.86                     & 98.62                     & \cellcolor[HTML]{DBFFDB}99.78                     & \cellcolor[HTML]{DBFFDB}99.95                     & \cellcolor[HTML]{DBFFDB}100.00                     & \cellcolor[HTML]{DBFFDB}100.00                     & \multicolumn{1}{c|}{98.25}                    & 98.60                     \\ \hline
\multicolumn{11}{l}{\cellcolor[HTML]{9AFF99}\textbf{Open-source}}                                                                                                                                                                                                           \\ 
\multicolumn{1}{l|}{Llama-3.3-70B-Instruct}    & \multicolumn{1}{c|}{70B}           & 88.91                     & 90.26                     & 98.07                     & 97.15                     & 98.86                     & 97.53                     & 97.85                     & \multicolumn{1}{c|}{93.72}                    & 95.29                     \\
\multicolumn{1}{l|}{Qwen3-32B}                 & \multicolumn{1}{c|}{32B}           & \cellcolor[HTML]{DBFFDB}99.47                     & 94.16                     & 97.73                     & 98.11                     & \cellcolor[HTML]{DBFFDB}100.00                     & \cellcolor[HTML]{DBFFDB}99.88                     & 98.60                     & \multicolumn{1}{c|}{\cellcolor[HTML]{DBFFDB}99.68}                    & 98.45                     \\
\multicolumn{1}{l|}{GLM-4-32B-0414}            & \multicolumn{1}{c|}{32B}           & 88.67                     & \cellcolor[HTML]{FFCCC9}68.18                     & 87.72                     & \cellcolor[HTML]{FFCCC9}67.67                     & 85.52                     & 86.58                     & \cellcolor[HTML]{FFCCC9}79.40                     & \multicolumn{1}{c|}{83.73}                    & 80.93                     \\
\multicolumn{1}{l|}{Devstral-Small-2507}       & \multicolumn{1}{c|}{24B}           & 82.92                     & 86.36                     & 93.45                     & \cellcolor[HTML]{FFCCC9}79.73                     & \cellcolor[HTML]{FFCCC9}70.10                     & 86.78                     & \cellcolor[HTML]{FFCCC9}78.70                     & \multicolumn{1}{c|}{82.53}                    & 82.57                     \\
\multicolumn{1}{l|}{Qwen3-14B}                 & \multicolumn{1}{c|}{14B}           & 96.23                     & 94.81                     & \cellcolor[HTML]{DBFFDB}99.45                     & \cellcolor[HTML]{DBFFDB}99.39                     & \cellcolor[HTML]{DBFFDB}99.00                     & \cellcolor[HTML]{DBFFDB}99.80                     & \cellcolor[HTML]{DBFFDB}99.80                     & \multicolumn{1}{c|}{98.62}                    & 98.39                     \\
\multicolumn{1}{l|}{GLM-4-9B-0414}             & \multicolumn{1}{c|}{9B}            & \cellcolor[HTML]{FFCCC9}76.06                     & \cellcolor[HTML]{FFCCC9}79.87                     & \cellcolor[HTML]{FFCCC9}78.08                     & 88.90                     & 93.35                     & 83.63                     & 91.50                     & \multicolumn{1}{c|}{\cellcolor[HTML]{FFCCC9}79.15}                    & 83.82                     \\
\multicolumn{1}{l|}{Qwen3-8B}                  & \multicolumn{1}{c|}{8B}            & 91.37                     & 92.86                     & \cellcolor[HTML]{DBFFDB}99.38                     & 98.07                     & 98.68                     & \cellcolor[HTML]{DBFFDB}99.40                     & 97.33                     & \multicolumn{1}{c|}{98.73}                    & 96.98                     \\
\multicolumn{1}{l|}{DeepSeek-R1-0528-Qwen3-8B} & \multicolumn{1}{c|}{8B}            & \cellcolor[HTML]{FFCCC9}67.43                     & \cellcolor[HTML]{FFCCC9}70.78                     & \cellcolor[HTML]{FFCCC9}64.58                     & 82.93                     & \cellcolor[HTML]{FFCCC9}67.88                     & \cellcolor[HTML]{FFCCC9}62.86                     & \cellcolor[HTML]{FFCCC9}64.88                     & \multicolumn{1}{c|}{\cellcolor[HTML]{FFCCC9}76.51}                    & \cellcolor[HTML]{FFCCC9}69.73                     \\
\multicolumn{1}{l|}{Llama-3.1-8B-Instruct} & \multicolumn{1}{c|}{8B}                & \cellcolor[HTML]{FFCCC9}29.93                     & \cellcolor[HTML]{FFCCC9}56.49                     & 90.63                     & \cellcolor[HTML]{FFCCC9}67.59                     & 85.98                     & \cellcolor[HTML]{FFCCC9}75.86                     & 86.90                     & \multicolumn{1}{c|}{\cellcolor[HTML]{FFCCC9}44.39}                    & \cellcolor[HTML]{FFCCC9}67.20                     \\ \hline
\end{tabular}
\caption{Task Completion Rate (TCR) of agentic LLMs on 8 tasks in SoMe. Avg. denotes the average TCR over 8 tasks. The values lower than 80 are highlighted in red and the values higher than 99 are highlighted in green.}
\label{tab:tcr}
\end{table*}

\subsection{Overall Results}
The experimental results of different agentic LLMs in social media agent tasks are demonstrated in Tables \ref{tab:res}-\ref{tab:tcr}. 

\textbf{Analysis of performance.} Based on Table \ref{tab:res}, we have the following observations: 
\begin{itemize}
    \item Both the current closed-source and open-source LLMs cannot handle social media agent tasks satisfactorily. 
    In most cases, current agentic LLMs typically fail to attain an evaluation score exceeding 70 in most tasks. This is particularly evident in tasks such as RED, SES, and MID, characterized by a high degree of openness in the information involved, where the majority of LLMs receive scores below 50.
    These results reveal that building social media agents is nontrivial, and more work is required to improve the performance and trustworthiness of social media agents.
    Our proposed SoMe provides a challenging yet meaningful testbed for future social media agents.

    \item Among API-based agentic LLMs, Gemini-2.5-Flash achieves the highest average performance score of 54.33. Furthermore, it demonstrates superior performance on individual tasks, including RED and SES with the scores of 54.92 and 44.87, respectively.
    Within the open-source agentic LLM category, Qwen3-32B attains the highest average score of 53.98. Moreover, it leads on key tasks such as UBP, MCR, and SMQ.

    \item Comparing the serial LLMs with different sizes, we could find that larger models typically outperform smaller models in social media agent tasks.
    For example, considering the average evaluation scores, Qwen3-32B outperforms Qwen3-14B by 2.2\%, Qwen3-14B outperforms Qwen3-8B by 3.8\%, and Llama-3.3-70B-Instruct outperforms Llama-3.1-8B-Instruct by 52.7\%.
    However, larger models typically have lower inference efficiency and face challenges regarding compatibility with local development.
    Especially for social media agent tasks that involve a large number of posts, user profiles, and even external knowledge, developing efficient and cost-effective agents is particularly crucial.

    \item Interestingly, while DeepSeek-R1-0528-Qwen3-8B significantly outperforms Qwen3-8B in reasoning tasks \cite{guo2025DeepSeek}, its capability declines when utilized as a social media agent.
    Compared to Qwen3-8B, DeepSeek-R1-0528-Qwen3-8B consistently experiences a decline in performance across 8 tasks by 56.14\%, 21.42\%, 41.07\%, 29.48\%, 35.88\%, 41.68\%, 43.30\%, and 31.95\%, respectively.
    These results indicate that agentic ability in social media tasks does not necessarily result from enhancing the reasoning skills of LLMs.
    Therefore, improving both the reasoning and agentic capabilities is crucial in the development of more powerful LLMs.

\end{itemize}

\textbf{Analysis of Task Completion Rate (TCR).} Based on Table \ref{tab:tcr}, we have the following observations: 
\begin{itemize}
    \item Since tasks in SoMe involve calling appropriate tools in correct orders, completing the task query is nontrivial.
    Some open-source LLMs, such as DeepSeek-R1-0528-Qwen3-8B and Llama-3.1-8B-Instruct, achieve relatively low TCR in handling social media agent tasks.
    Especially, Llama-3.1-8B-Instruct can only complete task queries with an average TCR of 65.76\%.
    These results indicate that improving the multi-round reasoning and tool-calling capability is still a challenge for some small open-source LLMs.

    \item Among all evaluated LLMs, Gemini-2.5-Flash, DeepSeek-V3, Qwen3-32B, and Qwen3-14B can handle tasks with extremely high average TCRs of 96.61\%, 98.60\%, 98.45\%, 98.39\%, respectively.
    These results show the outstanding agentic ability of these models, which can correctly understand and conduct tool-calling.
    Correspondingly, these models also achieve better performance on SoMe, as demonstrated in Table \ref{tab:res}.
    This implies that the ability to call tools and reason step-by-step is fundamental for social media agents.
\end{itemize}

\begin{figure}[ht]
\centering
\includegraphics[width=0.98\columnwidth]{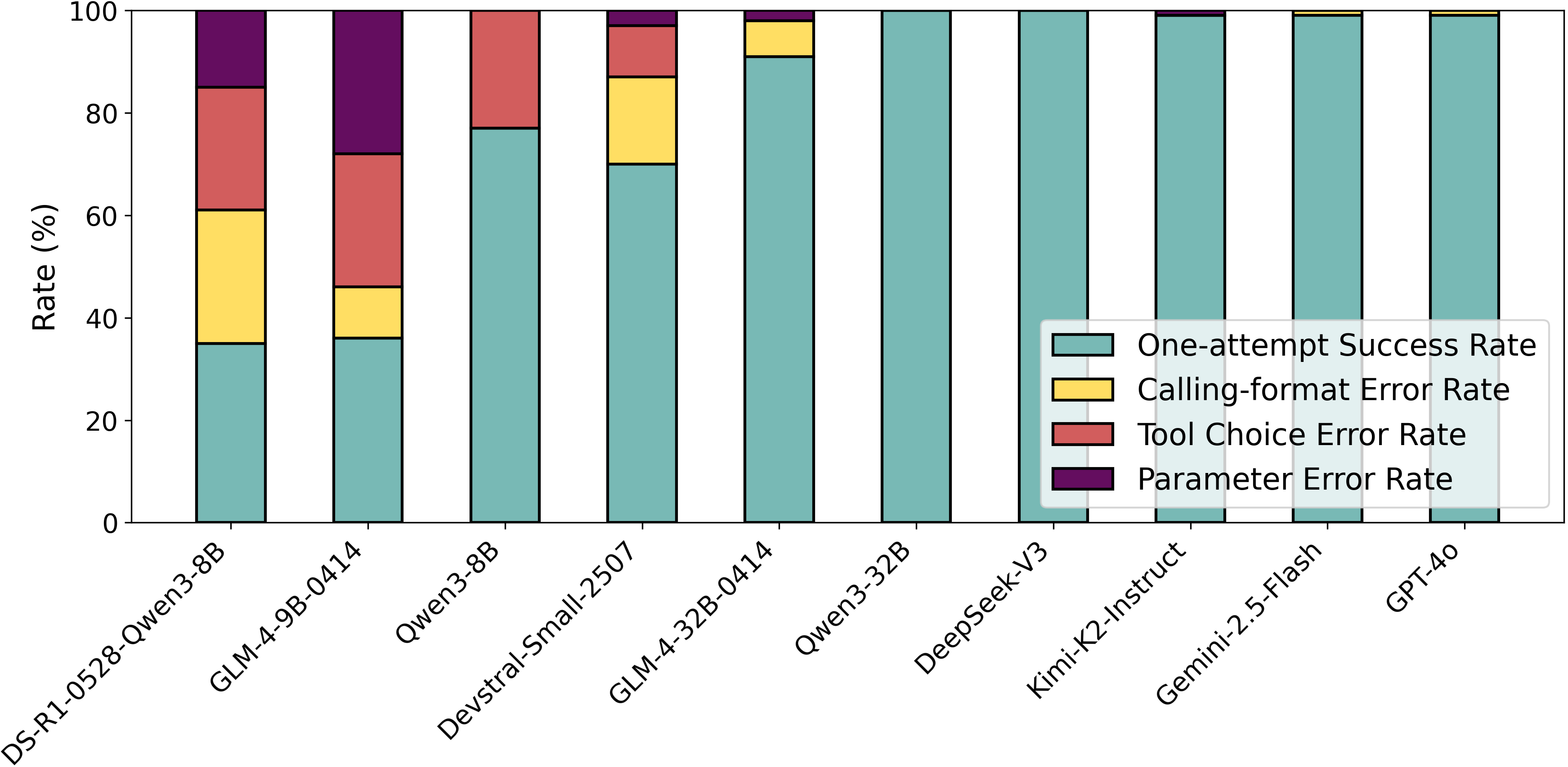}
\caption{One-attempt success rates and error rates of different agentic LLMs in tool-chain planning.}
\label{fig:toolchain}
\end{figure}

\begin{figure}[ht]
\centering
\includegraphics[width=0.98\columnwidth]{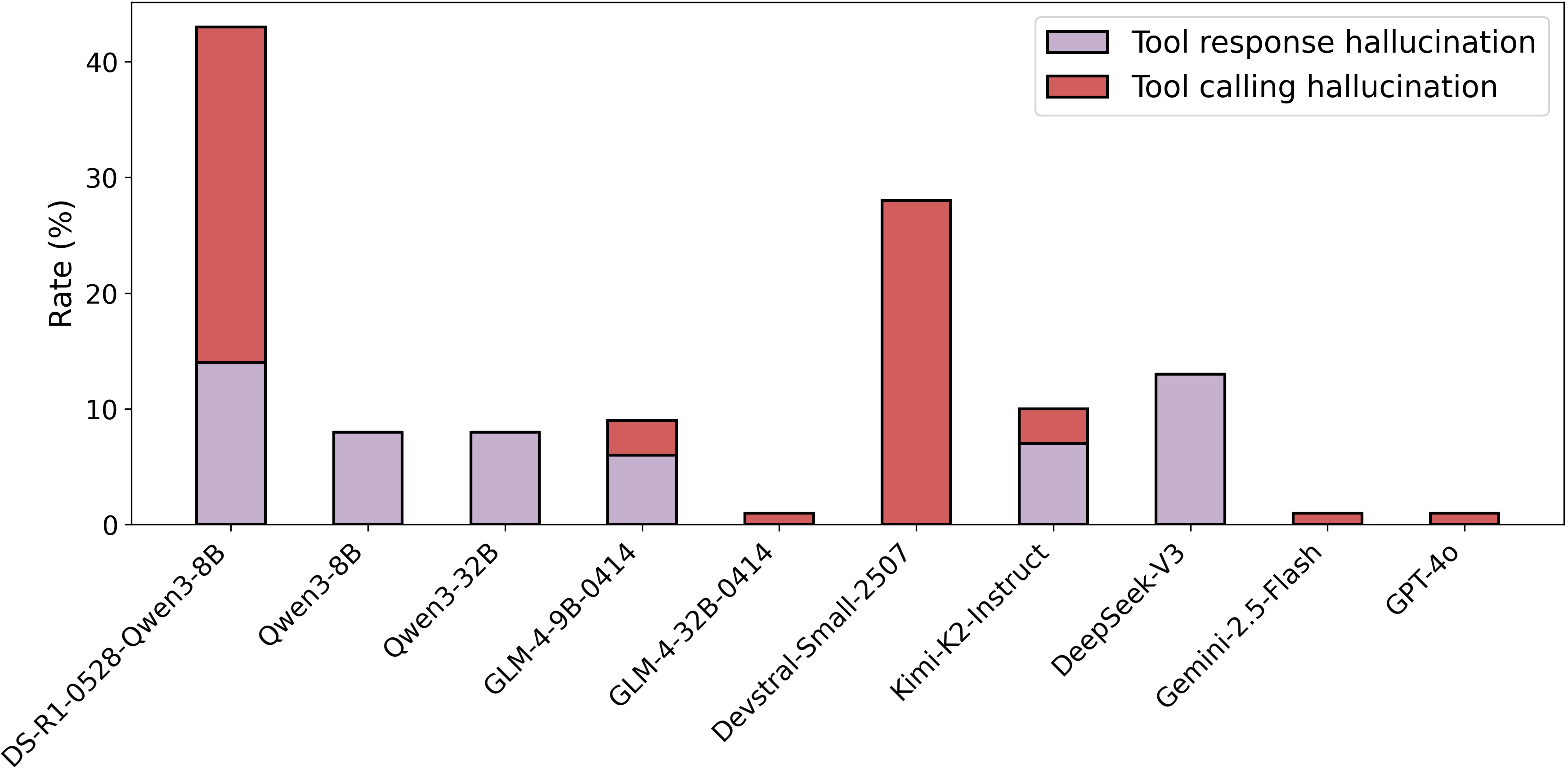}
\caption{Hallucination rates in tool-use of different agentic LLMs.}
\label{fig:hal}
\end{figure}

\subsection{Challenge Analysis}
 In this section, we conduct in-depth analyses of the typical errors made by mainstream agentic LLMs on SoMe, in order to identify the major challenges for future research.

\subsubsection{Errors in Planning Tool-chains.}
In complex social media tasks, the agents need to call tools multiple times in appropriate sequences to handle the task query. However, agents may encounter challenges in calling the tool-chains step-by-step.
Therefore, we further investigate the probabilities of social media agents successfully calling the tool-chain in a single attempt.
We report the One-attempt Success Rates (OSRs) in calling tool-chains on 100 samples of the real-time event detection task in Figure \ref{fig:toolchain}.
In the figure, we also outline the rates of different cases where the agents fail to call the next tool correctly, i.e., failing to generate the correct formats of tool-calling, choose the right tool, or provide the correct parameters for the tool.
Interestingly, we can observe that smaller models exhibit notably lower OSRs compared to larger models.
Qwen3-32B, DeepSeek-V3 (with 671B parameters), Kimi-K2-Instruct (with 1T parameters), Gemini-2.5-Flash, and GPT-4o achieve nearly 100\% success rates in accurately planning the tool-chains in one attempt.
GLM4-32B-0414 also significantly improves OSR compared to GLM4-9B-0414.
Small models generally struggle more in inferring the subsequent tool-calling based on past interactions.
%
%However, they possess inherent advantages in computational efficiency and compatibility.
%
Therefore, to develop superior social media agents, the tool-planning capability of smaller LLMs should be a focal point in future work.

\begin{figure}[ht]  
\centering
    \begin{minipage}{0.9\linewidth}
        \centerline{\includegraphics[width=\linewidth]{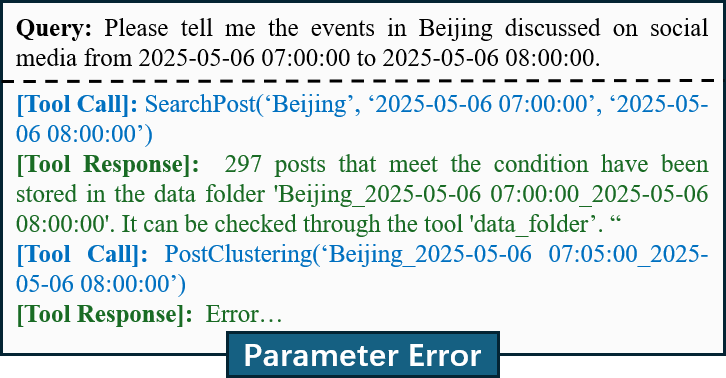}}
        \centerline{\footnotesize{(a) Kimi-K2-Instruct}}
    \end{minipage}
    \vfill
    \begin{minipage}{0.9\linewidth}
        \centerline{\includegraphics[width=\linewidth]{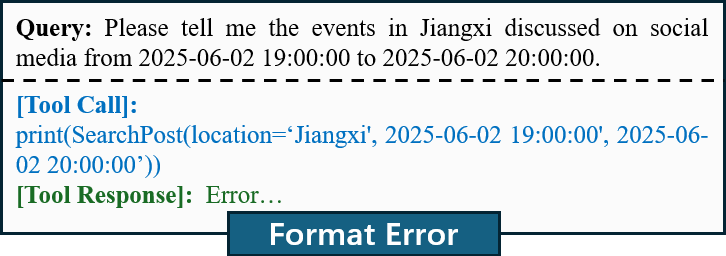}}
        \centerline{\footnotesize{(b) Gemini-2.5-Flash}}
    \end{minipage}  
    \vfill
    \caption{Typical errors made by Kimi-K2-Instruct and Gemini-2.5-Flash. Blue texts are generated by agents.}
    \label{fig:case}
\end{figure} 

\subsubsection{Hallucination in Tool-use}
In experiments, a common error scenario is observed where the agent fails to call the tool correctly and imagines the tool response or tool-calling format.
For example, in real-time event detection, the agent sometimes fails to call the tools and imagine the events satisfying the task query.
To further investigate the hallucination in tool-use, we design an experiment by intentionally altering the response of tools as errors and observing whether the agents will feign completing the task queries.
The experiments are conducted on 100 samples of the real-time event detection task.
As shown in Figure \ref{fig:hal}, most open-source LLMs (including DeepSeek-R1-0528-Qwen3-8B, Qwen3-8B, Qwen3-32B, GLM-4-9B-0414, Kimi-K2-Instruct, and DeepSeek-V3) suffer from the tool response hallucination, with the hallucination rates of 14\%, 8\%, 8\%, 6\%, 7\%, and 13\% individually.
Interestingly, this issue persists regardless of the model size increase, as even Kimi-K2-Instruct (with 1T parameters) and DeepSeek-V3 (with 671B parameters) are affected by the hallucination.
Moreover, the tool calling hallucination rates (which represent the rate of failing to call any tools due to the wrong format) of some open-source LLMs are also significantly high.
Specifically, DeepSeek-R1-0528-Qwen3-8B and Devstral-Small-2507 imagine the wrong format of tool-calling and fail to call any tools with a percentage of 29\% and 28\% individually.
However, it is noticeable that GLM-4-32B-0414 appears to address this prevalent limitation of open-source LLMs.
These results indicate that overcoming hallucination in tool-use is a significant challenge for current open-source models.

\subsubsection{Case Study}
We further point out some common errors made by top agentic LLMs in Figure \ref{fig:case}.
As shown in Figure \ref{fig:case}(a), Kimi-K2-Instruct sometimes fails to repeat the keywords in tool response or query, leading to tool parameter errors. 
As shown in Figure \ref{fig:case}(b), a common error made by Gemini-2.5-Flash is adding \textit{print($\cdot$)} outside the tool function.
These errors may reflect inherent flaws in training these state-of-the-art agentic LLMs, which should be considered in future work.

\section{Conclusion}
This paper introduces SoMe, a novel benchmark for evaluating LLM-based social media agents. SoMe comprises a diverse collection of 8 social media agent tasks, 9,164,284 posts, 6,591 user profiles, and 25,686 reports from 32 social media platforms and external websites, with 17,869 meticulously annotated task queries. Our experiments reveal that both the current closed-source and open-source agentic LLMs cannot handle social media agent tasks satisfactorily. Moreover, we identify several challenges encountered by the current agents through in-depth experimental analysis, in order to inspire future work. By providing a challenging benchmark, we hope to stimulate the development of advanced models capable of tackling the complexities of social media agent tasks.

\section{Acknowledgments}
This work is supported by the National Key Research and Development Program of China (No.2023YFC3310700), the Beijing Natural Science Foundation (JQ23018, L252032), and the National Natural Science Foundation of China (No.62276257).

\bibliography{aaai2026}

\appendix
\begin{figure}[ht]
\centering
\includegraphics[width=0.95\columnwidth]{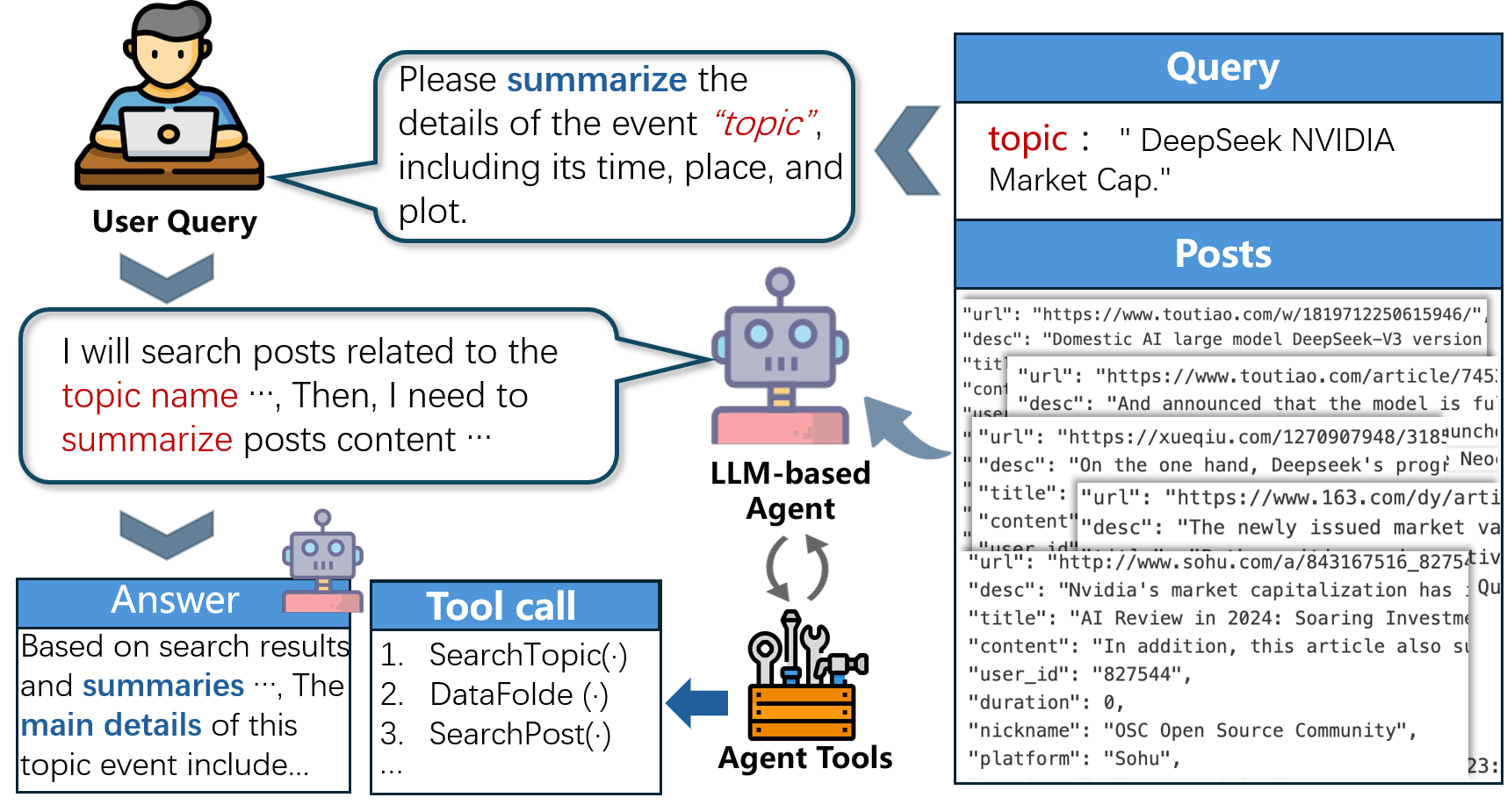}
\caption{An example for Streaming Event Summarization (SES).}
\label{exp1}
\end{figure}

\begin{figure}[ht]
\centering
\includegraphics[width=0.95\columnwidth]{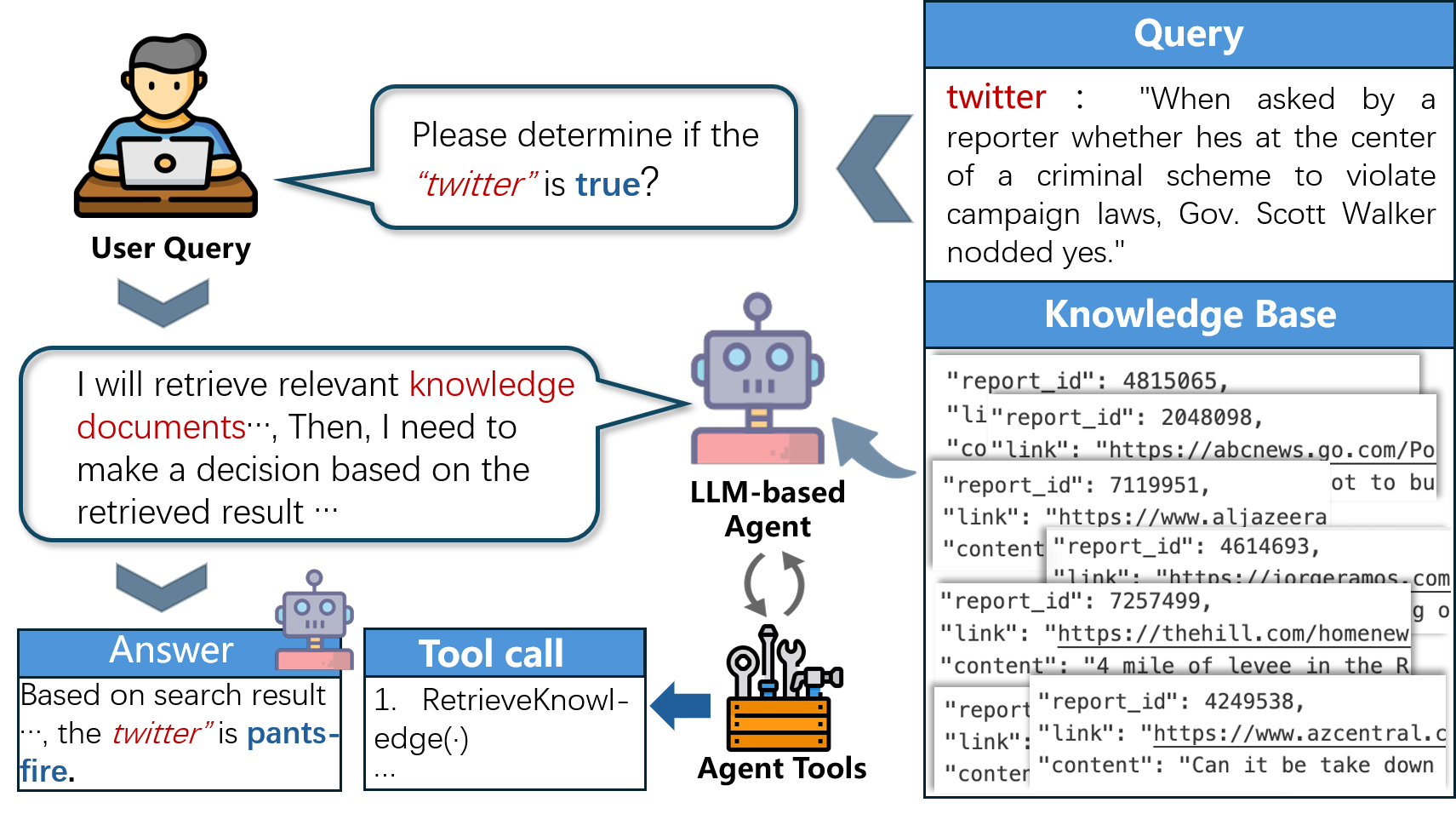}
\caption{An example for MisInformation Detection (MID).}
\label{exp2}
\end{figure}

\begin{figure}[ht]
\centering
\includegraphics[width=0.95\columnwidth]{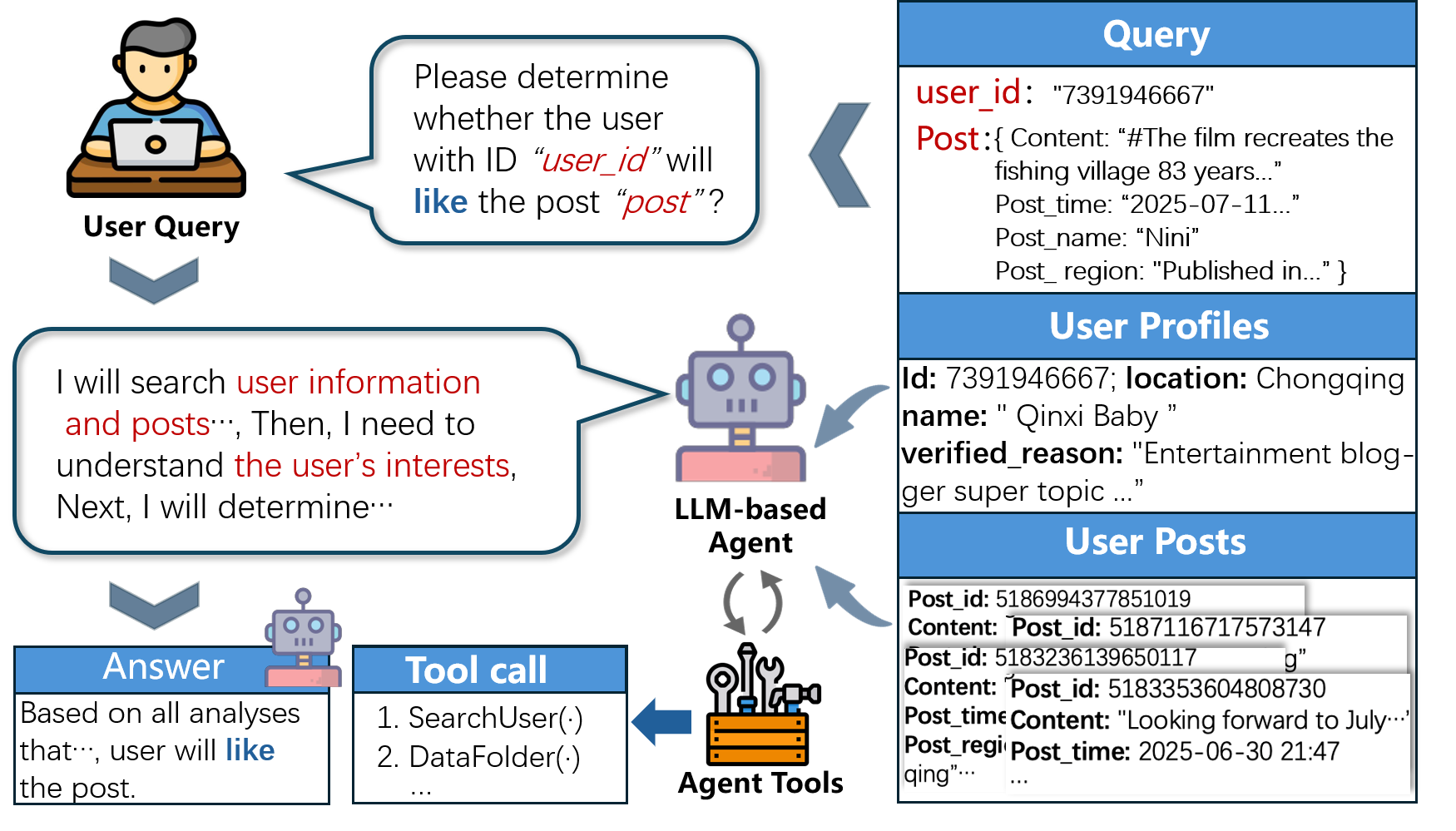}
\caption{An example for User Behavior Prediction (UBP).}
\label{exp3}
\end{figure}

\begin{figure}[ht]
\centering
\includegraphics[width=0.95\columnwidth]{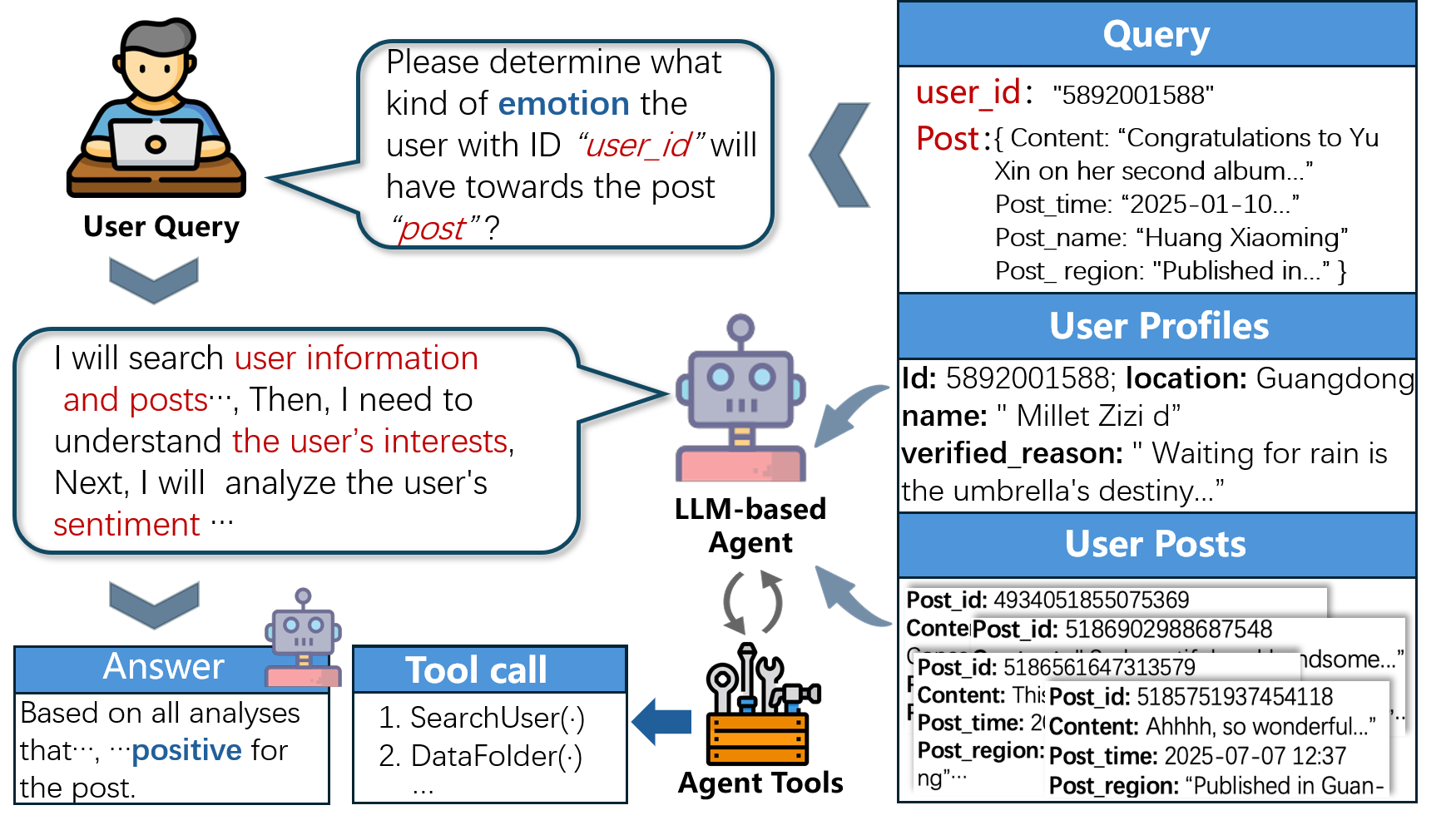}
\caption{An example for User Emotion Analysis (UEA).}
\label{exp4}
\end{figure}

\section{Task Definitions}
\begin{figure}[t]
\centering
\includegraphics[width=0.95\columnwidth]{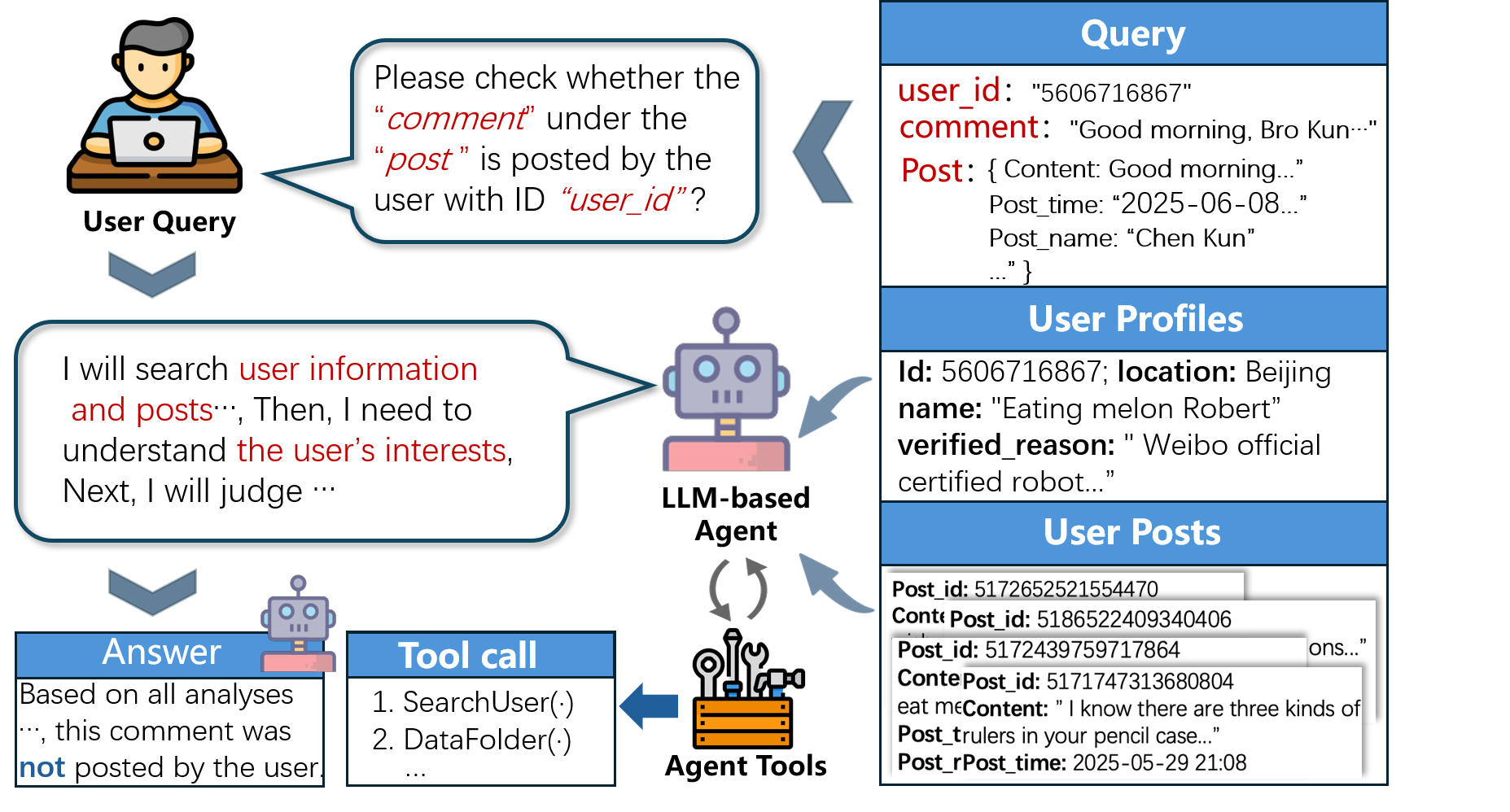}
\caption{An example for User Comment Simulation (UCS).}
\label{exp5}
\end{figure}

\begin{figure}[t]
\centering
\includegraphics[width=0.95\columnwidth]{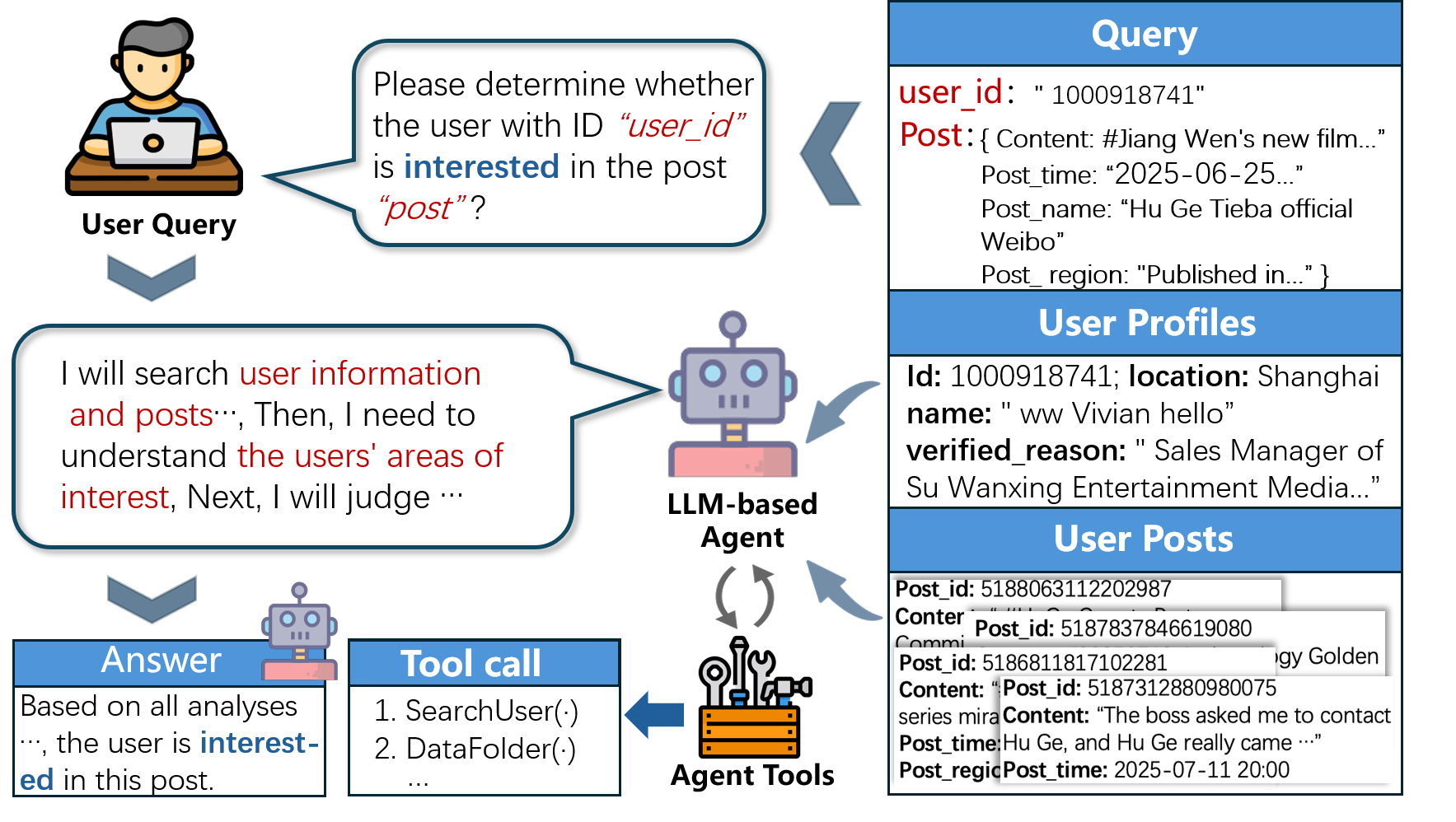}
\caption{An example for Media Content Recommendation (MCR).}
\label{exp6}
\end{figure}

\begin{figure}[t]
\centering
\includegraphics[width=0.95\columnwidth]{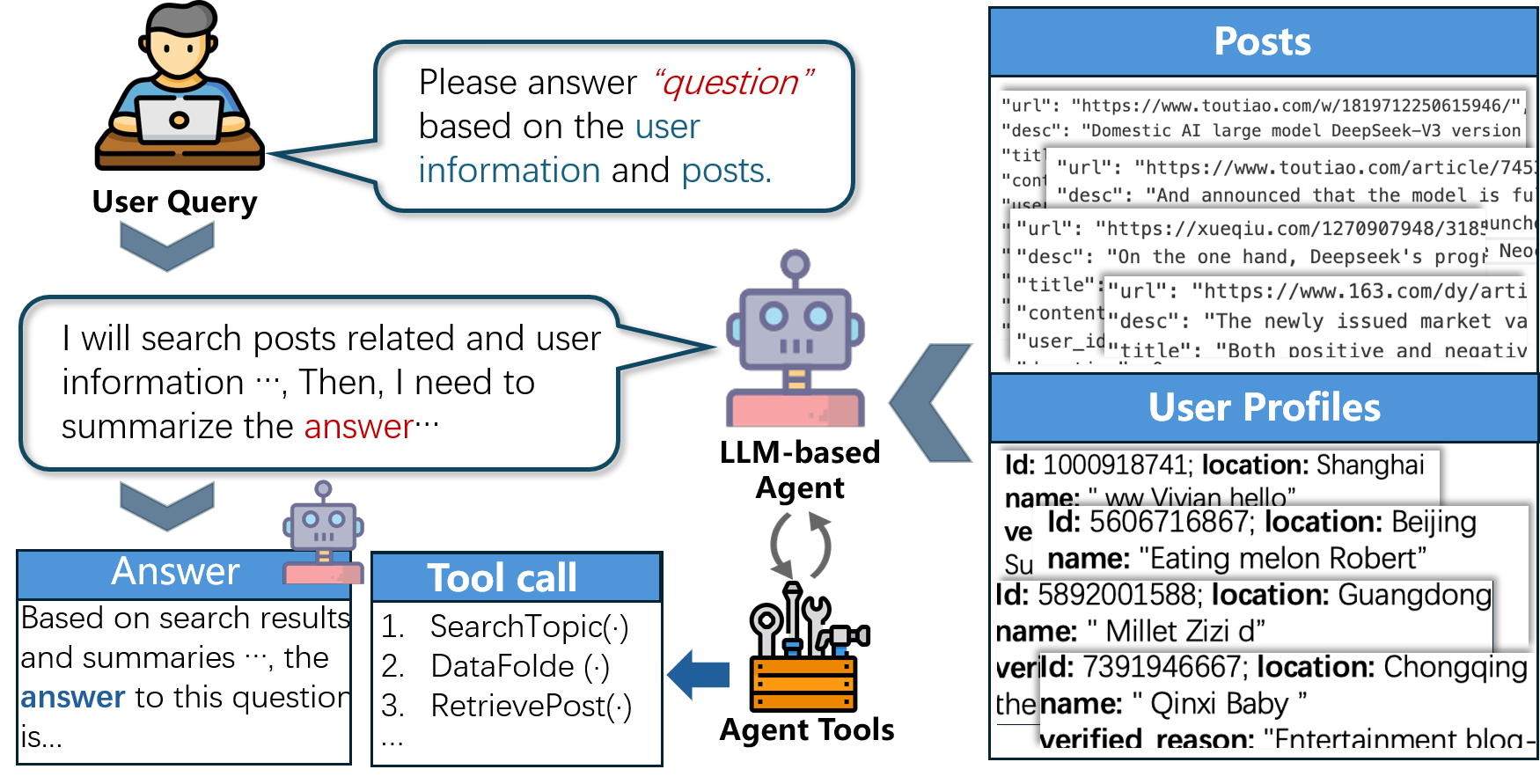}
\caption{An example for Social Media Question-answering (SMQ).}
\label{exp7}
\end{figure}

In this section, we provide further explanations of the comprised tasks in SoMe.
Except for the example for RED shown in Figure 1, the examples for other 7 tasks are shown in Figures \ref{exp1}-\ref{exp7}.
We explain these tasks in detail as following:
\begin{itemize}
    \item \textbf{Real-time Event Detection (RED).} As shown in Figure \ref{exp1}, this task aims to detect social events in real-time from a large volume of recent posts. Given a query specifying the target location and period, the agent will search the relevant posts from social media.
    These posts contain the location keywords but may not detail events that happened in the location.
    If the number of posts is small, the agent will directly read all posts and extract the events that match the specified location.
    If the posts are too many, the agent will cluster the posts and summarize every group.
    Subsequently, the agent will review all summaries to identify and extract valid events.

    \item \textbf{Streaming Event Summarization (SES).} This task aims to progressively summarize the details of a social event from continuously published posts. Given a query specifying the event topic, the agent will continuously search the related posts. 
    These posts contain extreme noise irrelevant to the event.
    Meanwhile, the agent will identify crucial details regarding the event and update the summarization of the event through a streaming process.

    \item \textbf{MisInformation Detection (MID).} As shown in Figure \ref{exp2}, this task aims to identify false, misleading, or inaccurate information within posts, utilizing support from external knowledge.
    Given a social media post, the agent will actively retrieve knowledge related to the key concept and claims in the post.
    By comparing materials, the agent will progressively find out any misinformation in the post.
    Finally, the agent will give a decision in \textit{pants-fire}, \textit{false}, \textit{barely-true}, \textit{half-true}, \textit{mostly-true}, and \textit{true} with a possible explanation.

    \item \textbf{User Behavior Prediction (UBP).} As shown in Figure \ref{exp3}, this task aims to predict the user interaction behaviors with specific posts.
    Given a query specifying the post, user, and interaction behavior, the agent first retrieves the user profile and historical posts, then analyzes their interests and past behavior.
    By comparing the user preferences with the content of the post, the agent will determine whether the user will like, comment, or repost the post.
    Finally,  the agent will give a decision in \textit{Yes}, \textit{No} with a possible explanation.

    \item \textbf{User Emotion Analysis (UEA).} As shown in Figure \ref{exp4}, this task aims to predict the emotions that emerge from users towards particular posts.
    Given a social media post and a user, the agent first retrieves the user profile and historical posts, then analyzes and extracts features such as user expression habits, areas of interest, and language emotions.
    By comparing the user preferences with the content of the post, the agent will find out the emotion that the user possibly has towards the post.
    Finally, the agent will give a decision in \textit{Positive}, \textit{Angry}, \textit{Sad}, \textit{Fear}, \textit{Surprise}, and \textit{Emotionless} with a possible explanation.

    \item \textbf{User Comment Simulation (UCS).} As shown in Figure \ref{exp5}, this task aims to predict the comments that users will make on given posts.
    Given a social media post, a comment under the post, and a user, the agent first retrieves the user profile and historical posts, then analyzes and extracts features such as user speaking style, common vocabulary, areas of interest, and expression habits.
    By comparing the user preferences with the content of the post, the agent will determine whether the comment on the post is posted by the user.
    Finally,  the agent will give a decision in \textit{Yes}, \textit{No} with a possible explanation.

    \item \textbf{Media Content Recommendation (MCR).} As shown in Figure \ref{exp6}, this task aims to recommend social media content that aligns with user preferences.
    Given a social media post, the agent first retrieves the user profile and historical posts, then analyze and extract areas or topics that users possible be interested in.
    By comparing the user preferences with the content of the post, the agent will determine whether the user is interested in this post.
    Finally,  the agent will give a decision in \textit{Yes}, \textit{No} with a possible explanation.

    \item \textbf{Social Media Question-answering (SMQ).} As shown in Figure \ref{exp7}, this task aims to answer questions regarding the public information available in posts and users.
    Given a specific query, the agent first determines whether the query is topic-related or user-related. Then, the agent will search for topic-related posts or user-related profiles based on the decision.
    By analyzing the search results, the agent will give the answer corresponding to the query.
    
\end{itemize}

\section{Tool Implementations}
In this section, we further explain the tools used in our social media agent framework.
\begin{itemize}
    \item \textbf{DataFolder(folder name, start idx, end idx).} This tool uses a dictionary to store data, with its display function saved in another dictionary with the same key. 
    The data is stored by other tools with the corresponding display function simultaneously.
    When calling \textbf{DataFolder(folder name, start idx, end idx)}, the data and its display function with the key of \textit{folder name} will be retrieved.
    Then, the data items from the start index \textit{start idx} to the end index \textit{end idx} will be organized into a readable string by the display function, and be returned to the agent.

    \item \textbf{SearchPost(location, start time, end time).} This tool search posts in the database published from \textit{start time} to \textit{end time} with keywords related to \textit{location} (e.g., sub-regions of \textit{location}). 
    The searched data will be stored in \textit{DataFolder} with a display function to show the posts.
    Finally, this tool will return the information about the data, including the folder name, to the agent.

    \item \textbf{SearchTopic(topic name).} This tool search posts in the database with keywords in \textit{topic name}. 
    The searched data will be stored in \textit{DataFolder} with a display function to show the posts.
    Finally, this tool will return the information about the data, including the folder name, to the agent.

    \item \textbf{SearchUser(uid).} This tool search the user of the id \textit{uid} in the database. 
    The published posts of the user will be stored in \textit{DataFolder} with a display function to show the posts.
    Finally, this tool will directly return the profile of the user and return the folder information of the posts.

    \item \textbf{RetrievePost(query, folder name, topk).} This tool retrieves semantically related posts with \textit{query} in the data folder of the key \textit{folder name}. 
    The retrieval is implemented by vector search, where the semantic vectors of \textit{query} and posts are embedded by Qwen3-Embedding-4B \cite{qwen3embedding}.
    Finally, this tool will return the top-k posts, where k equals \textit{topk}, organized by the display function to the agent.

    \item \textbf{RetrieveKnowledge(query, topk).} This tool retrieves semantically related reports with \textit{query} in the knowledge base. 
    The retrieval is also implemented by vector search, where the semantic vectors of \textit{query} and reports are embedded by Qwen3-Embedding-4B \cite{qwen3embedding}.
    Finally, this tool will return the top-k reports, where k equals \textit{topk}, organized by the built-in display function.

    \item \textbf{PostClustering(folder name).} This tool cluster posts in the data folder of the key \textit{folder name}, based on post embeddings encoded by Qwen3-Embedding-4B \cite{qwen3embedding}. 
    The clusters are grouped with a semantical similarity threshold of 0.5.
    Then, the post clusters will be stored in the data folder with a display function to show the clusters.
    Finally, this tool will return the information about the stored clusters, including the folder name, to the agent.

    \item \textbf{PostSummarization(folder name).} This tool summarizes clusters in the data folder of the key \textit{folder name}. 
    The posts in each cluster will be input into the backbone LLM of the social media agent, with event summarization prompts.
    Then, the summaries of clusters will be stored in the data folder with a display function to show the summaries.
    Finally, this tool will return the information about the stored summaries, including the folder name, to the agent.
\end{itemize}

\section{Dataset Annotation}
In this section, we introduce the annotation for every task in detail.

\subsection{Real-time Event Detection}
This task involves cross-platform social media data. To formulate task queries, we select three specific locations.
Each location is associated with 240  one-hour time periods over 10 days.
Each query is generated by the template \textit{``Please tell me the events in \{location\} discussed on social media from \{start time\} to \{end time\}."}.
Next, we conduct an automatic process to cluster posts related to each location and time period, filter out clusters irrelevant to any event, and generate an event summary for each cluster (comprising title, content, location, and time).
The clustering is conducted on daily data to improve the clustering performance with abundant data.
Following this, we conduct a manual annotation to check every event summary based on the source posts, rectify any inaccuracies, and remove invalid summaries.
After manual verification of every event summary, these summaries are linked to the task queries based on the location and time parameters in the queries.
After deleting queries without events (typically those from late-night local times), we finally obtain 568 task queries with ground truth event summaries.

\subsection{Streaming Event Summarization}
We collect 200 trending topics related to popular events from November 25th, 2024 to January 2nd, 2025, with massive related posts across social media platforms.
Each query is formulated by the template \textit{``Please summarize the details of the event\{topic\}, including its time, place, and plot."}.
To annotate the summary for each event topic, we utilize GPT-4o with the web search tool to generate the summary, given the keywords of the topic.
However, the searched content may be irrelevant to the target event and the summary may lack some key details.
Therefore, we conduct an intricate human-involved review by manually pointing out any inaccuracies in the current summary and refining the prompt to regenerate the summary.
After multiple rounds of improvements and rigorous quality verification, we finally obtain 154 event summaries in the dataset and remove unqualified events.

\subsection{Misinformation Detection}
For this task, we utilize the open-source LIAR-RAW and RAWFC datasets \cite{yang2022coarse}, where the ground truth is already annotated.
First, we merge the tweets and their labels in these two datasets.
In the source datasets, each tweet is associated with several reports to validate the information in the tweet.
We extract all these reports to form our knowledge base, which can be accessed by vector search.
Finally, we generate the task queries by the template \textit{``Please determine if the tweet ``\{tweet\}" is true."}.
We obtain 1,451 samples and 25,685 reports.

\subsection{User Behavior Prediction}
To annotate ground truth for this task, we directly utilize the crawled interaction behaviors in our data.
The task queries are structured by the template \textit{``Please determine whether the user with ID \{user\_id\} will \{interact\} with the post ``\{post\}"."}, where \textit{\{interact\}} could be like, repost, and comment.
The positive samples (with the label ``yes") are constructed using the real data of users.
To construct negative samples, for each positive sample, we randomly select another post that the same user has not interacted with the same interaction type.
This task does not involve LLM or manual annotation, and all samples are constructed based on the crawled real data.
Finally, we obtain 1,000 like samples, 1,000 repost samples, and 1,000 comment samples.

\subsection{User Emotion Analysis}
For this task, we utilize the crawled comments of users.
We input each comment on a post into GPT-4o to identify the user's emotion, categorizing it as positive, angry, sad, surprised, fearful, or emotionless.
To ensure accuracy, every comment undergoes three rounds of annotation to determine the predominant emotion through a voting mechanism.
Following this, we conduct a manual check to correct any inaccuracies in the labeled emotions and filter out inappropriate data.
Subsequently, we randomly delete some data of positive and emotionless to relieve class imbalance.
The task queries are constructed by the template \textit{``Please determine what kind of emotion the user with ID \{user\_id\} will have towards the post ``\{post\}"."}.
Finally, we obtain 800 positive samples, 184 angry samples, 550 sad samples, 337 surprised samples, 25 fearful samples, and 800 emotionless samples.

\subsection{User Comment Simulation}
This task also utilizes the crawled comment data.
The task queries are constructed by the template \textit{``Please check whether the comment \{comment\} under the post ``\{post\}" is posted by the user with ID \{user\_id\}."}.
The positive samples (with the label ``yes") are constructed using the real comments of users.
To construct hard negative samples, for each positive sample, we randomly select another comment by a different user under the same post as the negative sample.
We ensure the content of the negative comment is also different from the content of the positive comment.
Finally, we obtain 2,000 positive samples and 2,000 negative samples.

\subsection{Media Content Recommendation}
We utilize the crawled interaction data to annotate this task.
The task queries are constructed by the template \textit{``Please determine whether the user with ID \{user\_id\} is interested in the post ``\{post\}"."}.
The positive samples (with the label ``yes") are formed using posts that a user has interacted with.
Moreover, for each positive sample, we randomly select a post that the same user has not interacted with to serve as the negative sample. 
Furthermore, we guarantee that the content of the negative post differs from that of the positive post.
Finally, we obtain 2,000 positive samples and 2,000 negative samples.

\subsection{Social Meida Question-answering}
We extensively utilize the crawled topic data and user data to annotate this task.
We input the data of each topic or user with the published posts into GPT-4o to generate latent questions and answers for the social media data.
In this step, we generate 3,000 question-answer pairs for topics and 30,000 question-answer pairs.
Subsequently, we conduct a manual review to check the correctness of each pair and filter out pairs that have inaccurate answers, have vague questions, or can be answered without searching related social media data.
For the left pairs, we manually modify them to be more concise and accurate.
Finally, we obtain 1,000 question-answer pairs for topics and 1,000 question-answer pairs by removing pairs based on the type balance principle.
The questions are directly used as task queries.

\begin{figure*}[t]
\centering
\includegraphics[width=1.95\columnwidth]{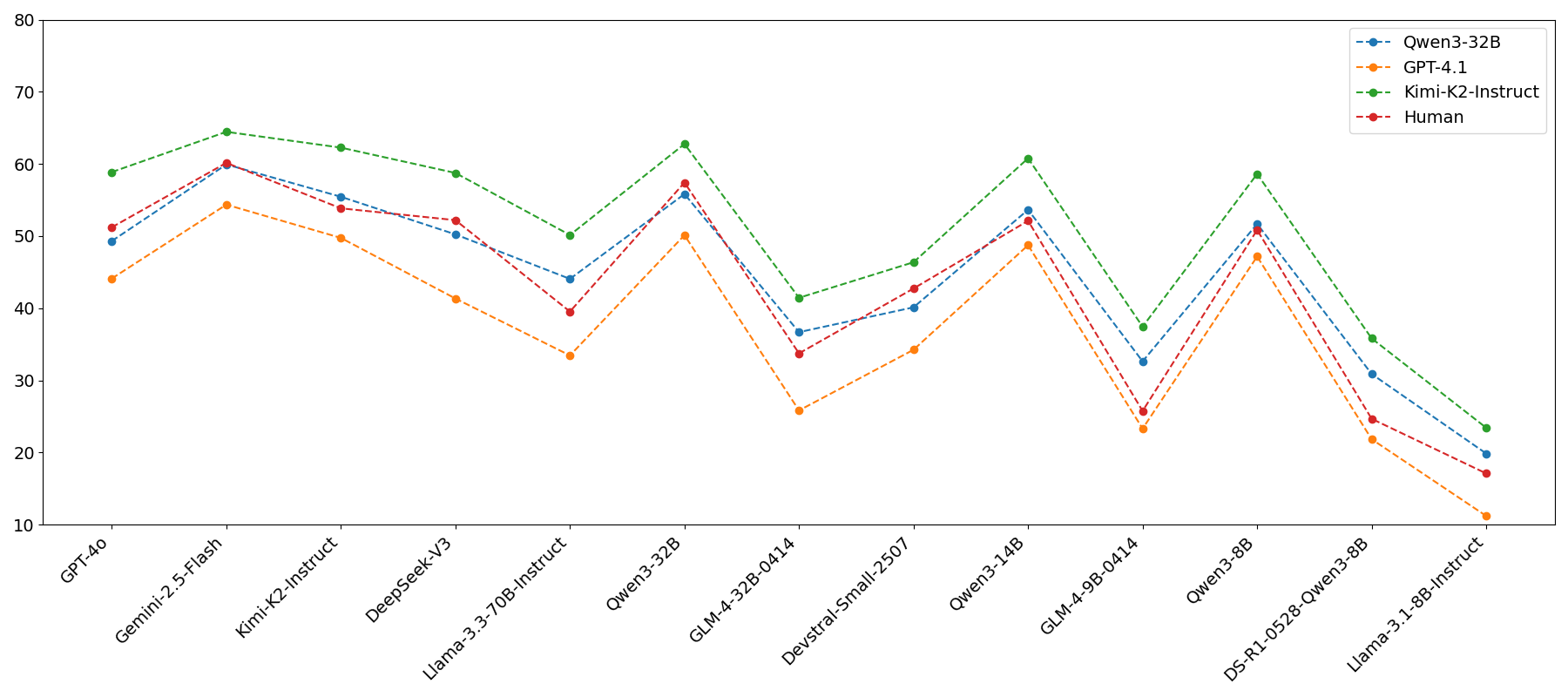}
\caption{Average scores over RED, SES, and SMQ evaluated by LLMs and human annotators.}
\label{fig:score}
\end{figure*}

\section{Details of Evaluation}
We develop a systematic approach to evaluate social media agents on SoMe.
While some tasks (MCR, UCS, UBP, MID, and UEA) can be formatted as multiple-choice tasks (e.g., select the user emotion from 6 choices), the generated answers by social media agents may contain explanations for the choices.
Therefore, we utilize LLM to extract the choice from the raw answer, of which the prompts are introduced as following.
After extracting the choices, we compute the accuracy of the answers based on ground truth labels.

Since the evaluation should be reproducible and accurate, we employ Qwen3-32B in the evaluation, as it is currently one of the most powerful open-source LLMs with an affordable memory requirement. We exclude closed-source models (such as Gemini and GPT-4o) since they may be deprecated and unavailable in the future.

\subsection{Answer Extraction Prompts}
Since the responses generated by the agents usually contain a significant amount of redundant information, we design task-specific prompts to effectively extract key answers using LLMs. 
Specifically, the Media Content Recommendation (MCR), User Comment Simulation (UCS), User Behavior Prediction (UBP), Misinformation Detection (MID), and User Emotion Analysis (UEA) tasks employ the LLM for answer extraction. 
We empirically find that the LLM can extract answers with 100\% accuracy.
The prompts for extracting answers are as follows.

\begin{table*}[!htb]
\centering
\begin{tcolorbox}[%`colback`=gray!10,
        colframe=black,
        width=\linewidth,
        arc=2mm, auto outer arc,
        title={Answer extraction prompt for Media Content Recommendation (MCR)}]		
Please carefully read the given text and extract the final judgment result from it to determine whether the user is likely to be interested in the post. The result should only be ``yes" or ``no" without any explanation or additional content.

Output format requirements:

- If the judgment result of the text content is positive, output: yes

- If the judgment result of the text content is negative, output: no

- If the judgment result of the text content is inconclusive or there is no judgment result, output: error

Note: Please strictly follow the above requirements when outputting the result.

Example input:

- ``Based on the information in the user search results, user 7391946667 primarily posted content about Jackson Yee's concerts, movies, and daily personal reflections and updates. From these specific posts, we can infer that this user may be interested in areas or topics such as concerts, movies, idol worship, and daily life."

Looking back at the five posts by this user, only the fifth mentions \#YiYangQianxi2025Concert\# and anticipates \#YiYangQianxiConcert\#, indicating that this user clearly has an interest in Jackson Yee's concerts.

Compare the user's interests with the given post content: ``With awe and curiosity, we once again step into the China Sky Eye. The starry sky will reveal the answers to our persistent exploration. ``Salute to science and technology workers, salute the scientific spirit, \#ScienceShineChina\#." This post primarily focuses on the ``China Sky Eye" and science and technology workers, and the topic doesn't match the user's interests. Ultimately, it's determined that the user may not be interested in this post.

Example output:

- ``No"

The given input text is ``\{raw answer\}"
\end{tcolorbox}
\end{table*}

\begin{table*}[!htb]
\centering
\begin{tcolorbox}[%`colback`=gray!10,
        colframe=black,
        width=\linewidth,
        arc=2mm, auto outer arc,
        title={Answer extraction prompt for User Comment Simulation (UCS)}]		
Please carefully read the given text and extract the final judgment result from it, determining whether the comment is likely to have been posted by the user. The result should only be ``Yes" or ``No" without any explanation or additional content.

Output format requirements:

- If the judgment result of the text is positive, output: Yes

- If the judgment result of the text is negative, output: No

- If the judgment result of the text is inconclusive, or there is no judgment result, output: Error

Note: Please strictly follow the above requirements when outputting the result.

Example Input:

- ``Based on analysis of the post content and comment style of user 7391946667 (Qinxi Baby), their comments primarily focus on topics related to Jackson Yee, daily greetings, and a small amount of event promotion. The comment ``It's my birthday today, please let me win a New Year's SVIP lottery" mentions birthdays and SVIP lottery content, but this user's past posts do not display similar lottery or birthday-related interactions. Furthermore, their language style (e.g., frequent use of simple greetings like ``good night" and ``good morning") is inconsistent with the casual tone of the comment." 

Furthermore, the mention of ``New Year's SVIP" in the comment has no direct connection to the user's recent posts. Their IP address is Chongqing, while the post they commented on was posted in Beijing, which also confuses them. Overall Verdict: No.

Example Output:

- ``No"

The given input text is ``\{raw answer\}"
\end{tcolorbox}
\end{table*}

\begin{table*}[!htb]
\centering
\begin{tcolorbox}[%`colback`=gray!10,
        colframe=black,
        width=\linewidth,
        arc=2mm, auto outer arc,
        title={Answer extraction prompt for User Behavior Prediction (UBP)}]		
Please carefully read the given text and extract the final judgment result from it to determine whether the user is likely to have performed the \{behavior\} action on the post. The result should only be ``Yes" or ``No" without any explanation or additional content.

Output format requirements:

- If the judgment result of the text content is positive, output: Yes

- If the judgment result of the text content is negative, output: No

- If the judgment result of the text content is indeterminate or no judgment result is found, output: Error

Note: Please strictly follow the above requirements when outputting the result.

Example Input:

- ``Based on an analysis of posts by user ``Ultraman and Little Monsters Watching the Stars" (ID: 5288580817), their primary activity is focused on celebrity fan interactions (especially Sun Qian), reposting Weibo posts, and a small amount of original content. They interact more with celebrity-related content, but less with other types of content (such as prayers for safety)".

The original post reads ``Praying for safety," and the poster is ``Li Xiaolu Super Lu," which is emotional content. Based on the user's historical behavior, their likes and interactions are primarily focused on celebrity fan activities, with less interaction on emotional content. Therefore, the overall judgment is:
No.

Example Output:

- ``No"

The given input text is ``\{raw answer\}"
\end{tcolorbox}
\end{table*}

\begin{table*}[!htb]
\centering
\begin{tcolorbox}[%`colback`=gray!10,
        colframe=black,
        width=\linewidth,
        arc=2mm, auto outer arc,
        title={Answer extraction prompt for MisInformation Detection (MID)}]		
Please read the provided text carefully and extract the final verdict indicating whether the given tweet contains any misinformation. The result must be one of the following: ``pants-fire", ``false", ``barely-true", ``half-true", ``mostly-true", or ``true". Do not provide any explanations or additional content.

You must explicitly extract and output the final verdict from the text.

Output format requirements:

- If the verdict is pants-fire, output: ``pants-fire".
    
- If the verdict is false, output: ``false".
    
- If the verdict is barely-true, output: ``barely-true".
    
- If the verdict is half-true, output: ``half-true".

- If the verdict is mostly-true, output: ``mostly-true".

- If the verdict is true, output: ``true".
    
- If the verdict is indeterminate or unclear, output: ``error".

You should always reponse to the user in English.

Note: Strictly follow the above output rules.

Example input:

-``Based on the retrieved documents, the statement `` Building a wall on the U.S.-Mexico border will take literally years" is accurate. Multiple sources indicate that constructing a border wall is a lengthy process, often taking several years due to logistical, financial, and political challenges. For instance, Document 1 mentions it could take more than three years, while Document 3 highlights experts' skepticism about completing it quickly".

Therefore, the statement is **mostly-true**.

Example output:

- ``mostly-true"

The given input text is ``\{raw answer\}"
\end{tcolorbox}
\end{table*}

\begin{table*}[!htb]
\centering
\begin{tcolorbox}[%`colback`=gray!10,
        colframe=black,
        width=\linewidth,
        arc=2mm, auto outer arc,
        title={Answer extraction prompt for User Emotion Analysis (UEA)}]		
Please carefully read the given text and extract the final judgment result, determining the user's emotion in response to the post. The result should only output one emotion; no explanation or additional content is required.

Output format requirements:

- If the emotion generated by the text content is judged to be positive, output: Positive

- If the emotion generated by the text content is judged to be angry, output: Angry

- If the emotion generated by the text content is judged to be sad, output: Sad

- If the emotion generated by the text content is judged to be fearful, output: Fear

- If the emotion generated by the text content is judged to be surprise, output: Surprise

- If the emotion generated by the text content is judged to be emotionless, output: Emotionless

- If the emotion generated by the text content is uncertain or no judgment result is given, output: Error

Note: Please strictly follow the above requirements when outputting the results.

Example input:

-- ``Based on the provided user information and post content, we can see that this user is a fan of Lu Han and frequently reposts and comments on posts about him. The user's profile and IP location indicate that they are located in Hong Kong and may be interested in entertainment news and celebrity updates from mainland China".

Combined with the post content, the user appears to be very interested in Lu Han's concerts and music, having reposted and commented on related posts numerous times. Furthermore, the user is also interested in Lu Han's personal life and updates, such as his selfies and daily life.

Thus, for a given post
\{Content: ``See you this weekend", Published Time: ``2024-11-19 23:24", Publisher: ``VueChen\_", Published Location: ``Published in Beijing", Reposts: 405411, Comments: 44368, Likes: 44368\}, users are likely to have **positive** emotions towards the post. This is because the post's content is about anticipation and wishes for the weekend, which may make users feel happy and excited. Furthermore, the poster, VueChen\_, may be a blogger or celebrity that users are interested in, further increasing their positive emotions towards the post.

Example output:

- ``Positive"

The given input text is ``\{raw answer\}"
\end{tcolorbox}
\end{table*}

\subsection{Answer Scoring Prompts}
In contrast, the Real-time Event Detection (RED), Streaming Event Summarization (SES), Social Media Question-answering (SMQ) tasks, which require stronger contextual understanding, are processed using the larger Qwen3-32B model. The prompts for extracting answers are as follows.

\begin{table*}[!htb]
\centering
\begin{tcolorbox}[%`colback`=gray!10,
        colframe=black,
        width=\linewidth,
        arc=2mm, auto outer arc,
        title={Answer scoring prompt for Real-time Event Detection (RED)},
        %breakable
        ]		
Please carefully read the reference answer ``\{ground truth\}" for the real-time event summary and rate the content of the real-time events on a scale of 0 to 5 based on the reference answer. The rating criteria include:

- Accuracy: Is the information in the content accurate? Do key details such as time, place, people, and events match the reference answer provided?

- Completeness: Does the content cover all important aspects mentioned in the reference answer? This includes but is not limited to the title, summary, time, and location.

- Relevance: Is the content relevant to the topic and content of the reference answer, and does it not include irrelevant information?

Please rate the content of the real-time event based on the three scoring dimensions and output a comprehensive score.

Accuracy scoring criteria are defined as follows:

- 5 points: Completely consistent with the reference answer, with no errors or omissions;

- 4 points: Mostly consistent with the reference answer, with a few minor errors or omissions;

- 3 points: Partially consistent with the reference answer, with some errors or omissions that do not affect overall understanding;

- 2 points: Limitedly consistent with the reference answer, with a number of errors or omissions that partially affect understanding;

- 1 point: Barely consistent with the reference answer, with serious errors or numerous omissions;

- 0 points: Completely inconsistent with the reference answer, with incorrect or blank information;

Completeness scoring criteria are defined as follows:

- 5 points: Completely consistent with the reference answer, covering all key time points, locations, and important events;

- 4 points: Mostly consistent with the reference answer, with a few minor omissions;

- 3 points: Partially matches the reference answer, but omits some key information, such as important events, time points, or locations.

- 2 points: Partially matches the reference answer, but omits a significant amount of key information.

- 1 point: Very limited information is mentioned, barely covering any key information in the reference answer.

- 0 point: Completely fails to cover the reference answer, with significant information missing or blank.

Relevance scoring criteria are defined as follows:

- 5 points: Highly relevant, focusing on key details mentioned in the reference answer.

- 4 points: Mostly relevant, the main content is related to the reference answer, but contains a small amount of irrelevant information.

- 3 points: Partially relevant, with some connection to the topic, but with significant inclusion of irrelevant content.

- 2 points: Weakly relevant, with most of the content deviating from the reference answer.

- 1 point: Almost irrelevant, with only a few words or concepts related to the reference answer.

- 0 points: Completely irrelevant, with no connection at all to the reference answer.

Please rate each dimension separately.

Output format requirements:

- If the scores for the three dimensions are: Accuracy: 2, Completeness: 2, Relevance: 2, then the output will be: Accuracy: 2, Completeness: 2, Relevance: 2

Note: Please strictly follow the above output requirements and do not add any explanations, clarifications, or superfluous content.

The given event summary is ``\{raw answer\}"
\end{tcolorbox}
\end{table*}

\begin{table*}[!htb]
\centering
\begin{tcolorbox}[%`colback`=gray!10,
        colframe=black,
        width=\linewidth,
        arc=2mm, auto outer arc,
        title={Answer scoring prompt for Streaming Event Summarization (SES)}]		
Please carefully read the reference answer for the topic event summary ``\{ground truth\}" and rate your summary based on the reference answer on a scale of 0 to 5. The scoring criteria include:

- Accuracy: Whether the time, events, and causal relationships in the reference answer are accurately reflected;

- Completeness: Whether the key time points and important events in the reference answer are fully covered.

Please rate your summary based on these two scoring criteria and output a comprehensive score.

The accuracy scoring criteria are defined as follows:

- 5 points: Completely consistent with the reference answer, with no errors or omissions;

- 4 points: Mostly consistent with the reference answer, with a few minor errors or omissions;

- 3 points: Partially consistent with the reference answer, with some errors or omissions that do not affect overall understanding;

- 2 points: Limitedly consistent with the reference answer, with a number of errors or omissions that affect partial understanding;

- 1 point: Barely consistent with the reference answer, with serious errors or numerous omissions;

- 0 points: Completely inconsistent with the reference answer, with incorrect or blank information;

Completeness scoring criteria are defined as follows:

- 5 points: Completely aligns with the reference answer, covering all key time points and important events;

- 4 points: Mostly aligns with the reference answer, with minor omissions;

- 3 points: Partially aligns with the reference answer, with omissions of some important events or time points;

- 2 points: Slightly aligns with the reference answer, with significant omissions of key information;

- 1 point: Barely aligns with the reference answer, with almost no key information covered;

- 0 points: Completely non-aligns with the reference answer, with significant omissions or blanks.

Please assign a score to each dimension separately.

Output Format Requirements:

- If the scores for two dimensions are: Accuracy: 2, Completeness: 2, then the output will be: Accuracy: 2, Completeness: 2

Note: Please output the results strictly according to the above requirements and do not add any explanations, clarifications, or superfluous content.

The given input text is ``\{raw answer\}"
\end{tcolorbox}
\end{table*}

\begin{table*}[!htb]
\centering
\begin{tcolorbox}[%`colback`=gray!10,
        colframe=black,
        width=\linewidth,
        arc=2mm, auto outer arc,
        title={Answer scoring prompt for Social Media Question-answering (SMQ)}]		
Please carefully read the reference answer ``\{ground truth\}" for the given question and rate your answer based on the reference answer on a scale of 0 to 5. The scoring criteria is:

- Accuracy: Whether the answer accurately reflects the key information in the reference answer;

Please rate your answer based on the scoring criteria.

The accuracy scoring scale is defined as follows:

- 5 points: Completely consistent with the reference answer, with no errors or omissions;

- 4 points: Mostly consistent with the reference answer, with a few minor errors or omissions;

- 3 points: Partially consistent with the reference answer, with some errors or omissions that do not affect overall understanding;

- 2 points: Limitedly consistent with the reference answer, with a number of errors or omissions that affect partial understanding;

- 1 point: Barely consistent with the reference answer, with serious errors or omissions;

- 0 points: Completely inconsistent with the reference answer, with incorrect or missing information.

Output Format Requirements:

- Directly output the accuracy score.

Note: Please output the results strictly according to the above requirements. Do not add any explanations, clarifications, or superfluous content.

The given input text is ``\{raw answer\}"
\end{tcolorbox}
\end{table*}

\subsection{Metric Computation}
After extracting answers for MCR, UCS, UBO, MID, and UEA, we compute the accuracy of agents as follows:
\begin{equation}
    ACC = P(\hat{y}=y),
\end{equation}
where $\hat{y}$ is the predicted answer and $y$ is the ground truth answer.
The ACC is computed on the whole test set for each task, separately.

After getting the LLM-based scores for RED, SES, and SMQ, we compute the average score over all criteria and all test samples for each task and agent separately.

\subsection{Comparison of LLM-based and Manual Evaluations}
As previously introduced, we employ Qwen3-32B to score the results generated by social media agents in the RED, SES, and SMQ tasks.
To ensure consistency with LLM assessments, we incorporate evaluations by other LLMs and human annotators.
Additionally, GPT-4.1 and Kimi-K2-Instruct are employed to evaluate all agents for the three tasks using the same prompts as previously mentioned.
Moreover, a random sample of 800 instances from the three tasks is selected, and 10 human annotators are engaged to assess the results produced by all agents.
The average scores across the three tasks are presented in Figure \ref{fig:score}.
We have the following observations: (1) The score disparities between human annotators and Qwen3-32B are generally consistent across various models, although human ratings exhibit greater discrimination, implying that the scores from Qwen3-32B are reasonably reliable.
(2) Kimi-K2-Instruct displays a tendency towards leniency, with over half of the models receiving scores above 50, while GPT-4.1 tends to assign lower scores.
For the sake of facilitating future reproducibility, we opt for Qwen3-32B in LLM-based evaluation since it is open-source and can be deployed locally with affordable costs.

\end{document}